\documentclass[lettersize,journal]{IEEEtran}

\usepackage{amsfonts}
\usepackage{amsmath} 
\usepackage{array}
\usepackage{algorithm}  
\usepackage{algorithmic}
\usepackage{verbatim} 
\usepackage{booktabs}
\ifCLASSOPTIONcompsoc
  \usepackage[caption=false,font=normalsize,labelfont=sf,textfont=sf]{subfig}
\else
  \usepackage[caption=false,font=footnotesize]{subfig}
\fi
\usepackage{textcomp}
\usepackage{stfloats}
\usepackage{float}
\usepackage{url}
\usepackage{graphicx}
\usepackage{amssymb}
\usepackage{color}

\usepackage{graphics}
\usepackage{epstopdf}
\usepackage{bbm}
\usepackage[T1]{fontenc}

\newtheorem{proposition}{\underline{Proposition}}

\newtheorem{remark}{\underline{Remark}}

\usepackage{balance}
\usepackage{hyperref}
\begin{document}

\title{Dual-Agent DRL based Cooperative Beamforming and Trajectory Optimization for UAV-ISAC Networks}

\author{Kaijian Yang,~\IEEEmembership{Student Member,~IEEE,} 
        Feng Wang,~\IEEEmembership{Member,~IEEE,} 
        Guojun Han,~\IEEEmembership{Senior Member,~IEEE,} \\
        Xin Wang,~\IEEEmembership{Fellow,~IEEE,} 
        and Vincent K. N. Lau,~\IEEEmembership{Fellow,~IEEE}

\thanks{K. Yang, F. Wang, and G. Han are with the School of Information Engineering, Guangdong University of Technology, Guangzhou 510006, China (e-mail: fengwang13@gdut.edu.cn).\emph{(Corresponding author: Feng Wang)} }
\thanks{X. Wang is with the Key Laboratory for Information Science of Electromagnetic Waves (MoE), College of Future Information Technology, Fudan University, Shanghai 200438, China (e-mail: xwang11@fudan.edu.cn).}
\thanks{V. K. N. Lau is with the Department of Electronic and Computer Engineering, The Hong Kong University of Science and Technology, Hong Kong, China (e-mail: eeknlau@ust.hk).}
\thanks{Copyright (c) 20xx IEEE. Personal use of this material is permitted. However, permission to use this material for any other purposes must be obtained from the IEEE by sending a request to pubs-permissions@ieee.org.}
}

\maketitle

\begin{abstract}
This paper considers a low-altitude unmanned aerial vehicle (UAV) network, where multiple ground base stations (BSs) provide integrated sensing and communication (ISAC) services for multiple communication UAVs (C-UAVs) and sensing UAVs (S-UAVs). With the objective of maximizing the average sum rate of C-UAVs over the entire flight episode, we jointly optimize the multi-BS multi-UAV ISAC beamforming and trajectory design, subject to the BS transmit power budget and the sensing signal-to-interference-plus-noise ratio (SINR) constraints of S-UAVs, as well as the flight mission and collision avoidance constraints of C-UAVs and S-UAVs. Based on the multi-agent deep reinforcement learning (MADRL) technique, we develop a dual-agent scheme, termed \emph{DeepDA-OPT}, to solve the resulting constrained Markov decision process (MDP). In the proposed DeepDA-OPT scheme, two agents, i.e., Agent-B and Agent-T, are specifically deployed for learning the multi-BS ISAC transmit beamformers and the C-UAV/S-UAV trajectories, respectively, where the optimal BS receive beamformers for the sensing echo signals of S-UAVs are obtained in closed form by leveraging the generalized Rayleigh quotient theorem. To mitigate the overestimation of value functions, both Agent-B and Agent-T are trained based on the twin-delayed deep deterministic (TD3) policy-gradient algorithm. In addition, a constraint-aware noise exploration policy is employed, and the reward function is designed to incorporate a sensing penalty and a flight urgency penalty, thereby guiding the training of Agent-B and Agent-T. Numerical results show that, compared with the baseline schemes, the proposed DeepDA-OPT scheme achieves a higher average sum rate of C-UAVs, while the sensing SINR constraints are empirically well satisfied and no collision violation is observed under the evaluated settings. Moreover, the proposed scheme exhibits more stable learning convergence across various network configurations.
\end{abstract}


\section{Introduction}
\IEEEPARstart{T}{he} rapid development of unmanned aerial vehicle (UAV) technology has catalyzed the explosive emergence of the low-altitude economy (LAE), including various mission-critical applications, such as intelligent logistics delivery, industrial inspection, and emergency response \cite{ref01, new01}. Typically, UAV deployment and operation require that aerial networks must simultaneously provide high-capacity broadband data communication and high-precision target sensing services. Intelligent network design also emphasizes information freshness and task utility \cite{Wu2026COMST}, as in federated digital twins optimizing age of information (AoI) for real-time healthcare \cite{Wu2025MNET}. By sharing spectral resources and hardware platforms, integrated sensing and communication (ISAC) has emerged as a key technique for 6G aerial networks to enable the physical integration of data transmission and target detection, thereby flexibly accommodating the size, weight, and power constraints of UAVs \cite{ref02, ref04,ref14}.

In complex UAV environments, single base station (BS) ISAC systems face inherent physical bottlenecks in coverage and sensing resolution due to line-of-sight (LoS) blockages and severe cellular cross-interference. Furthermore, a single BS sensing perspective is highly susceptible to UAV radar cross section (RCS) fluctuations. To overcome these fundamental limitations, cooperative multi-BS ISAC architectures have emerged as a promising solution \cite{new12}. By exploiting the joint beamforming of distributed antenna arrays and collaborative signal processing, cooperative multi-BS architectures can effectively eliminate spatial blind spots and exploit macroscopic diversity to enhance positioning accuracy, thereby achieving an enhanced ISAC performance \cite{new06, new07}. Note that the cooperative multi-BS ISAC system must provide high-capacity data communicate service for UAVs and track highly mobile UAVs using radar sensing beams, while simultaneously adjusting dynamic trajectories via command and control links. Therefore, it calls for joint optimization of continuous UAV trajectory and multi-BS ISAC beamforming designs in a highly dynamic LAE airspace, which is a challenging and high-dimensional problem \cite{ref07, ref09}.

\subsection{Related Work}
Currently, extensive research has been devoted to the integration of UAVs and ISAC systems. Depending on the role of UAVs, existing studies can be broadly categorized into two main paradigms. The first paradigm is known as UAV-assisted ISAC services, where UAVs act as aerial platforms to provide communication and sensing services. For instance, \cite{work4} maximized target localization accuracy by jointly designing the 3D trajectory and transmit covariance of a single UAV. To improve coverage, \cite{work7} explored a multi-UAV cooperative architecture to enhance overall sensing performance under communication quality of service (QoS) constraints by jointly optimizing transmit beamformers and trajectories. \cite{work8} jointly designed UAV trajectories and ground-user association to balance downlink rates and the Cram\'er--Rao bound (CRB) for ground-user tracking. \cite{reviewer_jsac_iscpt} jointly optimized association/scheduling, beamforming, and UAV trajectories in a multi-UAV-assisted integrated sensing, communication, and power transfer (ISCPT) network. To circumvent the high computational complexity inherent in traditional optimization methods, several learning-based UAV communication and ISAC schemes have been developed in \cite{ref12,work13,ref23}. For RIS-assisted UAV communications, \cite{Huang2025TMC} accelerated energy-efficient trajectory learning via multithreading and federated reinforcement learning (RL), instantiated as FedSAC/FedPPO.

The limitation of the first paradigm is that it is difficult to apply directly to LAE scenarios where UAVs serve as both communication users and sensing targets. The second paradigm leverages ground BSs to provide ISAC services for UAVs to meet the practical demands of the LAE. Early research primarily focused on single-BS scenarios. Specifically, \cite{work1} proposed a predictive beamforming scheme based on the extended Kalman filter for continuous UAV tracking. Beyond beam alignment, \cite{work2} combined deterministic sensing and Gaussian information signals for bistatic UAV detection, optimizing transmit beamforming to maximize the minimum detection probability under communication constraints. \cite{work3} used artificial noise (AN) for eavesdropping-UAV tracking and jamming, jointly optimizing near-field BS beamforming and communication-UAV trajectories to maximize the instantaneous secrecy rate.

To address spatial cross-interference in multi-UAV setups, \cite{work5} considered UAVs as both users and targets and designed a target-specific beamforming scheme to achieve synergistic performance improvements. As network scale expands, the coverage limitations of a single BS become prominent. Consequently, the works \cite{ref05,ref06} explored multi-BS cooperative networks. For multi-BS architectures, \cite{ref05} maximized the downlink communication sum rate under sensing constraints by jointly optimizing cooperative beamformers and UAV trajectories. \cite{ref06} jointly optimized transmit/receive beamforming and UAV trajectories for cooperative BS communication and UAV sensing. In vehicular edge networks, \cite{Ning2021TITS} jointly optimized offloading and caching to reduce delay. For wireless federated vehicular platooning, \cite{Wu2025TON} combined analytical resource allocation with multi-agent reinforcement learning (MARL) for vehicle selection, considering AoI and federated model drift. In UAV-assisted collaborative vehicular learning, \cite{Ding2026TAAS} combined coalition formation and UAV resource-allocation games with proximal policy optimization (PPO) to address knowledge diversity and catastrophic forgetting. For mobile autonomous systems, \cite{Ding2026MNET} developed sparse layerwise federated unlearning with concurrent learning and unlearning for efficient data removal.

Model-based UAV-ISAC design widely uses iterative optimization; DRL/MARL has also been introduced for sequential control. \cite{ref15} used single-agent DRL for joint GBS beamforming and UAV trajectory control in LAE. \cite{work10} introduced meta-DRL for energy-efficient joint beamforming and trajectory control in dynamic environments. More recently, dual-agent SA-TTD3 \cite{ref18} and HR-MAPPO \cite{ref19} addressed UAV-enabled ISAC beamforming/trajectory optimization. \cite{work11} used MARL for distributed UAV-swarm trajectory and power control. Table~\ref{tab:related_work_comparison} compares representative UAV-ISAC studies. 

\begin{table*}[t]
\centering
\caption{Comparison with representative UAV-ISAC studies.}
\label{tab:related_work_comparison}
\scriptsize
\renewcommand{\arraystretch}{1.0}
\setlength{\tabcolsep}{2pt}
\begin{tabular*}{\textwidth}{@{\extracolsep{\fill}}>{\centering\arraybackslash}p{0.043\textwidth}>{\centering\arraybackslash}p{0.075\textwidth}>{\centering\arraybackslash}p{0.081\textwidth}>{\centering\arraybackslash}p{0.088\textwidth}>{\centering\arraybackslash}p{0.071\textwidth}>{\centering\arraybackslash}p{0.125\textwidth}>{\centering\arraybackslash}p{0.190\textwidth}>{\centering\arraybackslash}p{0.190\textwidth}@{}}
\toprule
\textbf{Ref.} & \textbf{Multi-BS} & \textbf{UAV comm. user} & \textbf{UAV sensing target} & \textbf{Multi-UAV} & \textbf{Performance metric} &\textbf{Design variables} & \textbf{Method} \\
\midrule
\cite{work8} & $\times$ & $\times$ & $\times$ & {$\checkmark$} & {Weighted rate/CRB} & {UAV traj., GU assoc.} & {EKF/AO/SCA} \\
{\cite{reviewer_jsac_iscpt}} & {$\times$} & {$\times$} & {$\times$} & {$\checkmark$} & {WPT efficiency} & {Assoc./sched., BF, traj.} & {Two-stage opt.} \\
{\cite{work13}} & {$\times$} & {$\times$} & {$\times$} & {$\checkmark$} & {Min. weighted SE} & {Assoc., power, traj.} & {SAC/MASAC} \\
{\cite{work1}} & {$\times$} & {$\checkmark$} & {$\checkmark$} & {$\times$} & {Outage capacity} & {Pred. traj., sensing ratio, SNR} & {EKF/SCA/AO} \\
{\cite{work2}} & {$\times$} & {$\checkmark$} & {$\checkmark$} & {$\checkmark$} & {Min. detection prob.} & {Tx BF, sensing covariance} & {NP detector/SDR/SCA} \\
{\cite{work3}} & {$\times$} & {$\checkmark$} & {$\checkmark$} & {$\checkmark$} & {Secrecy rate} & {Tx BF/AN, C-UAV traj.} & {EKF/AO/SCA} \\
{\cite{work5}} & {$\times$} & {$\checkmark$} & {$\checkmark$} & {$\checkmark$} & {Assoc./comm.} & {Specific BF, ID assoc.} & {DIA/EKF} \\

{\cite{ref05}} & {$\checkmark$} & {$\checkmark$} & {$\times$} & {$\checkmark$} & {Avg. sum rate} & {Tx BF, traj., assoc.} & {AO/SCA/SDR} \\
{\cite{ref06}} & {$\checkmark$} & {$\checkmark$} & {$\checkmark$} & {$\checkmark$} & {WSR} & {Tx/Rx BF, traj.} & {AO/SDR/SCA} \\
{\cite{ref15}} & {$\times$} & {$\checkmark$} & {$\times$} & {$\checkmark$} & {Exp. sum rate} & {Tx BF, UAV traj.} & {DeepLSC (DRL)} \\
{\cite{ref18}} & {$\times$} & {$\times$} & {$\times$} & {$\times$} & {Comm. sum rate} & {BF, UAV traj., STAR-RIS} & {SA-TD3} \\
{\cite{ref19}} & {$\times$} & {$\times$} & {$\times$} & {$\checkmark$} & {Sum data rate} & {BF, UAV traj., assoc.} & {HR-MAPPO} \\
{Ours} & {$\checkmark$} & {$\checkmark$} & {$\checkmark$} & {$\checkmark$} & {Avg. sum rate} & {Tx/Rx BF, C/S-UAV traj.} & {Closed-form Rx + dual-TD3} \\
\bottomrule
\end{tabular*}
\end{table*}

\subsection{Main Contributions}
 In this work, we consider a multi-BS ISAC system with multiple communication UAVs (C-UAVs) and sensing UAVs (S-UAVs). Subject to the per-slot sensing signal-to-interference-plus-noise ratio (SINR) constraints of S-UAVs, we aim to maximize the downlink average sum rate of C-UAVs over the entire flight episode by jointly optimizing BS ISAC beamforming and UAV trajectory design. In addition, the C-UAV/S-UAV flight mission and collision avoidance constraints, as well as the per-BS maximum transmit power constraints, are included. The resulting joint beamforming and trajectory design problem is a challenging non-convex optimization problem, whose high-dimensional and heterogeneous decision structure motivates a learning-based solution. The main contributions are summarized as follows. 
\begin{itemize}
    \item First, leveraging the MADRL technique, we develop a structure-aware ``analyze-then-learn'' framework, termed DeepDA-OPT, for the joint optimization of BS transmit beamforming and C-UAV/S-UAV trajectory planning. For given transmit beamformers and UAV trajectories, the sensing receive beamformers are obtained in closed form from the generalized Rayleigh quotient theorem and are thereby removed from the learned action space. The remaining heterogeneous decision mappings are learned by Agent-B for high-dimensional multi-BS transmit beamforming and Agent-T for UAV kinematic control, respectively. This factorization does not reduce the physical control degrees of freedom; rather, it avoids requiring a single policy to simultaneously represent the heterogeneous beamforming and mobility actions.
    
    \item Second, both Agent-B and Agent-T are trained using the twin-delayed deep deterministic (TD3) policy-gradient algorithm to mitigate action-value overestimation. To promote flight-mission completion under DRL exploration uncertainties, an environment-side constrained action mapping rectifies the trajectory action when the remaining flight time is insufficient to complete the mission. The collision-avoidance and sensing constraints are handled through soft reward penalties; hence, their satisfaction is assessed empirically rather than presented as a strict deterministic guarantee.
    
    \item Third, a customized reward function is designed in DeepDA-OPT to prevent the constrained exploration strategy from interfering with the policy learning of Agent-B and Agent-T. In particular, the sensing penalty is evaluated based on the multi-BS closed-form receive beamformers, which provides unbiased feedback to guide Agent-B toward superior beamforming policies. Moreover, a flight urgency penalty is designed for C-UAVs/S-UAVs to prevent Agent-T from overly relying on the constrained exploration mechanism, thereby guiding it to learn smooth trajectory policies.
    
    \item Finally, extensive simulations compare DeepDA-OPT with the conventional AO-based scheme \cite{ref06}, SA-TD3 \cite{ref14}, SA-DDPG \cite{ref23}, DeepDA-MRC (maximum ratio combining), and the SA-TD3-MRC baseline. The latter two MRC-based variants, together with their analytical-receiver counterparts, form a 2$\times$2 ablation study that separates the effect of the analytical sensing receiver from that of the dual-agent policy factorization. The results show that the analytical receiver improves the achieved sensing--communication operating point, whereas the dual-agent factorization provides a complementary improvement in learning efficiency and run-to-run stability.
\end{itemize}

 The rest of the paper is organized as follows. Section \ref{sec:system_model} introduces the cooperative UAV-ISAC system model. Section \ref{sec:problem_formulation} formulates the joint optimization problem and analytically derives the optimal closed-form receive beamformers via the generalized Rayleigh quotient theorem. Section \ref{sec:mdp_reformulation} establishes the dual-agent MDP reformulation. Section \ref{sec:proposed_scheme} details the proposed DeepDA-OPT scheme. Section \ref{sec:simulation} presents the numerical results, and Section \ref{sec:conclusion} concludes this paper.

\section{System Model}
\label{sec:system_model}
In this work, we consider a multi-BS multi-UAV ISAC system consisting of a BS set $\mathcal{M}\triangleq\{1,...,M\}$ of $M$ cooperative BSs, a C-UAV set $\mathcal{K}_c\triangleq\{1,...,K_c\}$ of $K_c$ C-UAVs, and a S-UAV set $\mathcal{K}_s\triangleq\{1,...,K_s\}$ of $K_s$ S-UAVs, where the S-UAVs act as targets to be sensed by the BSs and simultaneously the BSs communicate with each C-UAV for data exchange. Each BS $m\in{\cal M}$ is assumed to be equipped with a uniform planar array (UPA) of size $S = S_W \times S_L$, where the antenna element spacing is $d = \frac{\lambda}{2}$ and $\lambda$ denotes the carrier wavelength. % \cite{ref20}

\begin{figure}%[t]
 \centering
\includegraphics[width=3.0in]{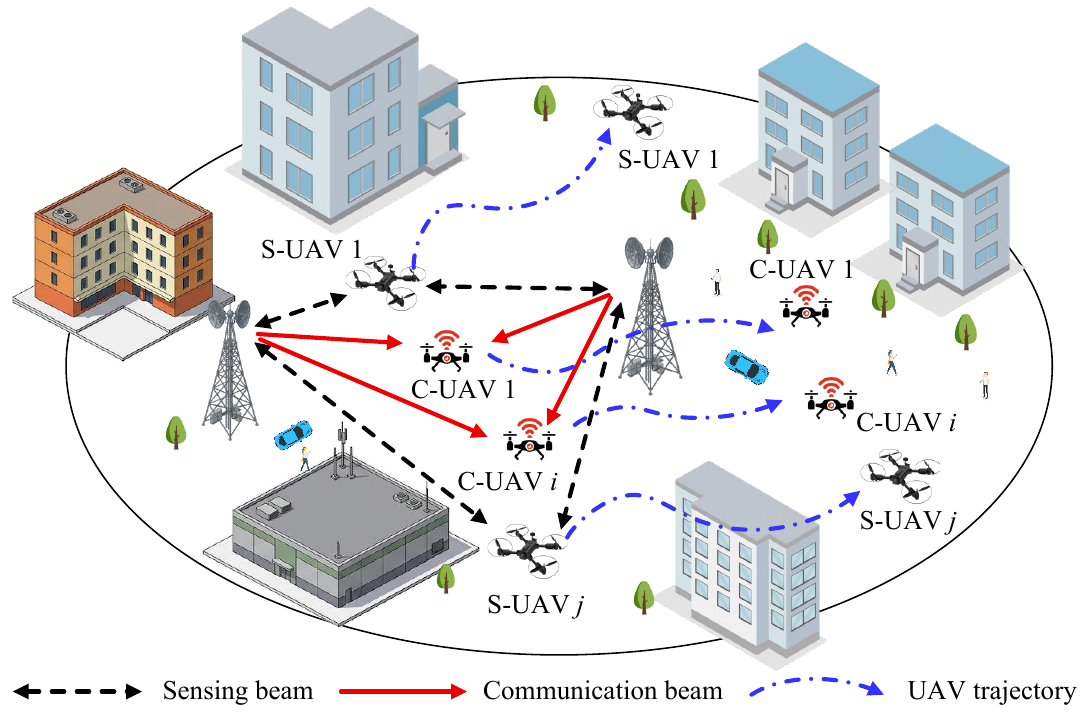}
\caption{Multi-BS multi-UAV ISAC system model.}
 \label{fig:fig1}
\end{figure}

\subsection{C-UAV/S-UAV Trajectory Planning Model}
 For convenience, the total C-UAV/S-UAV flight episode $T$ is divided into a set ${\cal N}\triangleq\{0,1,...,N-1\}$ of $N$ time slots, where the duration of each time slot is fixed as $\Delta t = T/N$. The trajectories of each C-UAV $i\in{\cal K}_c$ and S-UAV $j\in{\cal K}_s$ are denoted as $N$-length sequences $\{\mathbf{q}_i^c[n]|n=0,1,...,N-1\}$ and $\{\mathbf{q}_j^s[n]|n=0,1,...,N-1\}$, respectively, where $\mathbf{q}^c_i[n] \triangleq (x^c_i[n], y^c_i[n])$ and $\mathbf{q}^s_j[n] \triangleq (x^s_j[n], y^s_j[n])$ denote the locations of C-UAV $i\in{\cal K}_c$ and $j\in{\cal K}_s$ at slot $n$, respectively, and the flying altitude is fixed as $H>0$. We assume that the UAV current location are available at the beginning of each slot as state information for channel and beamforming optimization.
 The initial and final locations of C-UAVs/S-UAVs are given by
 \begin{subequations}\label{eq:trajectory1}
     \begin{align}
   &\mathbf{q}_i^c[0] = \mathbf{\bar{q}}^c_i, \quad \mathbf{q}_i^c[N] = \mathbf{\tilde{q}}^c_i, \quad \forall i \in \mathcal{K}_c \\
   &\mathbf{q}_j^s[0] = \mathbf{\bar{q}}^s_j, \quad \mathbf{q}_j^s[N] = \mathbf{\tilde{q}}^s_j, \quad \forall j \in \mathcal{K}_s,
     \end{align}
 \end{subequations}
where $\mathbf{\bar{q}}^c_i$ ($\mathbf{\bar{q}}^s_j$) and $\mathbf{\tilde{q}}^c_i$ ($\mathbf{\tilde{q}}^s_j$) denote the initial location and final location of C-UAV $i$ (S-UAV $j$), respectively. Denote by $v_i^c[n]$ and $v_j^s[n]$ the velocities of C-UAV $i$ and S-UAV $j$ during time slot $n\in{\cal N}$, respectively. The UAV flight distance constraints are given by
\begin{subequations}\label{eq:trajectory2}
    \begin{align}
    & \|\mathbf{q}_i^c[n] - \mathbf{q}_i^c[n-1]\|=v_i^c[n]\Delta t \leq V_{\max} \Delta t, \forall i \in \mathcal{K}_c \\
    &\|\mathbf{q}_j^s[n] - \mathbf{q}_j^s[n-1]\|=v_j^s[n]\Delta t \leq V_{\max} \Delta t, \forall j \in \mathcal{K}_s, 
    \end{align}
\end{subequations}
where $V_{\max}$ denotes the UAV maximum velocity. In addition, to ensure airspace safety and avoid collision among C-UAVs/S-UAVs, we impose the collision avoidance constraints at each slot $n\in{\cal N}$ as 
\begin{subequations}\label{eq:trajectory3}
    \begin{align}
    & \|\mathbf{q}_i^c[n] - \mathbf{q}_{i'}^c[n]\| \geq d_{\min}, \quad \forall i \neq i' \in \mathcal{K}_c \\
    & \|\mathbf{q}_j^s[n] - \mathbf{q}_{j'}^s[n]\| \geq d_{\min}, \quad \forall j \neq j' \in \mathcal{K}_s \\
    & \|\mathbf{q}_i^c[n] - \mathbf{q}_j^s[n]\| \geq d_{\min}, \quad \forall i \in \mathcal{K}_c, j \in \mathcal{K}_s,   
    \end{align}
\end{subequations}
where $d_{\min}$ is the minimum allowed safety distance between two UAVs. 

\subsection{ISAC Signal Model}
At each time slot $n\in{\cal N}$, the ISAC signal transmitted by each BS $m\in{\cal M}$ is expressed as
\begin{equation}
\mathbf{x}_m[n] = \sum_{i \in \mathcal{K}_c} \mathbf{w}_{m,i}^c[n] s_{m,i}^c[n] + \sum_{j \in \mathcal{K}_s} \mathbf{w}_{m,j}^s[n] s_{m,j}^s[n],
\label{eq:transmitted_signal}
\end{equation}
where $s_{m,i}^c[n]\in\mathbb{C}$ and $s_{m,j}^s[n]\in\mathbb{C}$ represent the independent data symbols transmitted by BS $m$ to C-UAV $i\in{\cal K}_c$ and S-UAV $j\in{\cal K}_s$, respectively, and the vectors $\mathbf{w}_{m,i}^c[n] \in \mathbb{C}^{S \times 1}$ and $\mathbf{w}_{m,j}^s[n] \in \mathbb{C}^{S \times 1}$ are the corresponding ISAC transmit beamforming weights at BS $m$. Without loss of generality, we assume both $s_{m,i}^c[n]$ and $s_{m,j}^s[n]$ are independent random variables with zero mean and unit variance. Therefore, we have the following ISAC transmit power constraints
\begin{align}\label{eq:ISAC_power}
\sum_{i \in \mathcal{K}_c} \|\mathbf{w}^c_{m,i}[n]\|^2 + \sum_{j \in \mathcal{K}_s} \|\mathbf{w}^s_{m,j}[n]\|^2 \leq P_{\max}, \forall n\in{\cal N},
\end{align}
where $P_{\max}$ denotes the per-slot transmit power budget at each BS $m\in{\cal M}$.

Note that the low-altitude airspace is typically dominated by LoS links, and non-line-of-sight components are neglected for facilitating the ISAC beamforming designs. Therefore, the downlink channel vector from BS $m\in{\cal M}$ to C-UAV $i\in{\cal K}_c$ and S-UAV $j\in{\cal K}_s$ at time slot $n\in{\cal N}$ are denoted as 
\begin{subequations}
\begin{align}
   \mathbf{h}^c_{m,i}[n] &= \sqrt{\beta^c_{m,i}[n]} \tilde{\mathbf{h}}^c_{m,i}[n] \\
      \mathbf{h}^s_{m,j}[n] &= \sqrt{\beta^s_{m,j}[n]} \tilde{\mathbf{h}}^s_{m,j}[n],
\end{align} 
\end{subequations}
respectively, where $\beta^c_{m,i}[n] \triangleq L_0 ( \|\mathbf{q}^c_i[n] - \mathbf{v}_m\|^2 + H^2)^{-1}$ and $\beta^s_{m,j}[n] \triangleq L_0 ( \|\mathbf{q}^s_j[n] - \mathbf{v}_m\|^2 + H^2)^{-1}$ with $L_0$ representing the channel power gain at the reference distance of one meter and $\mathbf{v}_m\triangleq (x_m, y_m)$ being the location of BS $m$; $\tilde{\mathbf{h}}^c_{m,i}[n]$ and $\tilde{\mathbf{h}}^s_{m,j}[n]$ denote the steering vectors from BS $m$ to C-UAV $i$ and S-UAV $j$, respectively. It then follows that
\begin{subequations}\label{eq:steering_vector}
\begin{align}
\tilde{\mathbf{h}}^c_{m,i}[n] &= \mathbf{a}(\theta^c_{m,i}[n], \phi^c_{m,i}[n]) \otimes \mathbf{b}(\theta^c_{m,i}[n], \phi^c_{m,i}[n])\\
\tilde{\mathbf{h}}^s_{m,j}[n] &= \mathbf{a}(\theta^s_{m,j}[n], \phi^s_{m,j}[n]) \otimes \mathbf{b}(\theta^s_{m,j}[n], \phi^s_{m,j}[n]),
\end{align}
\end{subequations}
where $\otimes$ denotes Kronecker product; $(\theta^c_{m,i}[n],\phi^c_{m,i}[n])$ and $(\theta^s_{m,j}[n],\phi^s_{m,j}[n])$ denote the elevation/azimuth angle tuples from BS $m$ to C-UAV $i$ and to S-UAV $j$ at slot $n\in{\cal N}$, respectively. Specifically, the elevation/azimuth angles of C-UAVs/S-UAVs at each time slot $n\in{\cal N}$ can be computed by $\theta^c_{m,i}[n] = \arctan( \|\mathbf{q}_i^c[n]-\mathbf{v}_m\|/{H})$, $\phi^c_{m,i}[n] = \text{arctan2}( y_i^c[n] - y_m, x_i^c[n] - x_m)$,
$\theta^s_{m,j}[n] = \arctan( {\|\mathbf{q}_j^s[n]-\mathbf{v}_m\|}/{H})$, and $\phi^s_{m,j}[n] = \text{arctan2}( y_j^s[n] - y_m, x_j^s[n] - x_m)$\cite{new19}. 

In \eqref{eq:steering_vector}, the terms $\mathbf{a}(\theta, \phi)$ and $\mathbf{b}(\theta, \phi)$ denote the response vectors of the BS UPA along the width and length dimensions, respectively, and we have
 \begin{subequations}
 \begin{align}
&\mathbf{a}(\theta, \phi) = [1, e^{-j\frac{2\pi d}{\lambda} \tilde{\theta}}, \dots, e^{-j\frac{2\pi d}{\lambda} (S_W - 1) \tilde{\theta}}]^T \\
&\mathbf{b}(\theta, \phi) = [1, e^{-j\frac{2\pi d}{\lambda} \hat{\theta}}, \dots, e^{-j\frac{2\pi d}{\lambda} (S_L - 1) \hat{\theta}}]^T,
\end{align}
\end{subequations}
where $\tilde{\theta}\triangleq\sin \theta \cos \phi$ and $\hat{\theta}\triangleq\sin \theta \sin \phi$.

\subsection{Communication Model}
Denote by $y_i[n]\in\mathbb{C}$ the baseband signal received by C-UAV $i\in{\cal K}_c$ at each slot $n\in{\cal N}$, which is modeled as 
\begin{equation}
\begin{split}
y_i[n] &= \underbrace{\sum_{m \in \mathcal{M}} \mathbf{h}_{m,i}^{cH}[n] \mathbf{w}_{m,i}^c[n] s_{m,i}^c[n]}_{\text{Desired comm. signal}} \\
&\quad + \underbrace{\sum_{m \in \mathcal{M}} \sum_{i' \in \mathcal{K}_c \setminus \{i\}} \mathbf{h}_{m,i}^{cH}[n] \mathbf{w}_{m,i'}^c[n] s_{m,i'}^c[n]}_{\text{Multi-user interference (MUI)}} \\
&\quad + \underbrace{\sum_{m \in \mathcal{M}} \sum_{j \in \mathcal{K}_s} \mathbf{h}_{m,i}^{cH}[n] \mathbf{w}_{m,j}^s[n] s_{m,j}^s[n]}_{\text{Sensing interference (SI)}} + \underbrace{n_i[n]}_{\text{noise}},
\end{split}
\label{eq:received_signal_model}
\end{equation}
where $\mathbf{h}^c_{m,i}\in\mathbb{C}^{S\times 1}$ is the downlink channel vector from BS $m\in{\cal M}$ to C-UAV $i\in{\cal K}_c$, and $n_i[n] \sim \mathcal{CN}(0, \sigma_i^2)$ denotes the additive white Gaussian noise (AWGN) at C-UAV $i$. Based on \eqref{eq:received_signal_model}, the SINR of C-UAV $i$ at time slot $n$ is given by 
\begin{equation}
\gamma_i^c[n] = \frac{\sum_{m \in \mathcal{M}} \left| \mathbf{h}_{m,i}^{cH}[n] \mathbf{w}_{m,i}^c[n] \right|^2}
{ I_{i}^{\text{MUI}}[n] + I_{i}^{\text{SI}}[n] + \sigma_i^2 },~\forall i\in{\cal K}_c,
\label{eq:sinr_c}
\end{equation}
where the powers of multi-user interference (MUI) $I_{i}^{\text{MUI}}[n]$ and sensing interference (SI) $I_{i}^{\text{SI}}[n]$ are calculated respectively as 
$I_{i}^{\text{MUI}}[n] = \sum_{m \in \mathcal{M}} \sum_{i' \in \mathcal{K}_c \setminus \{i\}} \big| \mathbf{h}_{m,i}^{cH}[n] \mathbf{w}_{m,i'}^c[n] \big|^2$ and $I_{i}^{\text{SI}}[n] = \sum_{m \in \mathcal{M}} \sum_{j \in \mathcal{K}_s} \big| \mathbf{h}_{m,i}^{cH}[n] \mathbf{w}_{m,j}^s[n] \big|^2$. Therefore, the downlink achievable average sum rate of the $K_c$ C-UAVs at each slot $n\in{\cal N}$ is expressed as
\begin{equation}
R[n] = \sum_{i \in \mathcal{K}_c} R_i[n] = \sum_{i \in \mathcal{K}_c} \log_2\left(1 + \gamma_i^c[n]\right).
\label{eq:weighted_sum_rate}
\end{equation}

\subsection{Sensing Model}
 At time slot $n$, denote by $\mathbf{y}_{r,m}[n]\in\mathbb{C}^{S\times 1}$ the echo signal vector received by BS $m\in{\cal M}$, which is given by
\begin{align}
\mathbf{y}_{r,m}[n] &= \underbrace{\sum_{i \in \mathcal{K}_c} \xi^c_{m,i} \mathbf{A}^c_{m,i}[n] \mathbf{x}_m[n]}_{\text{Echo signal from all C-UAVs}} + \underbrace{\sum_{j \in \mathcal{K}_s} \xi^s_{m,j} \mathbf{A}^s_{m,j}[n] \mathbf{x}_m[n]}_{\text{Echo signal from all S-UAVs}} \notag \\
&+ \underbrace{\sum_{l \in \mathcal{M} \setminus \{m\}} \sqrt{\eta_{l,m}} \mathbf{H}^{\text{IBI}}_{l,m}[n] \mathbf{x}_l[n]}_{\text{Inter-BS-interference}} \notag \\
&+ \underbrace{\sqrt{\zeta_m} \mathbf{H}^\text{SI}_m[n] \mathbf{x}_m[n]}_{\text{Self-interference}} + \mathbf{n}_{r,m}[n],
\label{eq:received_sensing_signal}
\end{align}
where $\mathbf{A}^c_{m,i}[n] \triangleq \mathbf{h}^c_{m,i}[n]\mathbf{h}^{cH}_{m,i}[n]$ and $\mathbf{A}^s_{m,j}[n] \triangleq \mathbf{h}^s_{m,j}[n]\mathbf{h}^{sH}_{m,j}[n]$; $\xi^c_{m,i}\sim {\cal CN}(0,\sigma_t^2)$ and $\xi^s_{m,j}\sim {\cal CN}(0,\sigma_t^2)$ denote the RCS values of C-UAV $i$ and S-UAV $j$, respectively; $\mathbf{H}^{\text{SI}}_{m}\in \mathbb{C}^{S \times S}$ and $\mathbf{H}^{\text{IBI}}_{l,m}\in \mathbb{C}^{S \times S}$ denote the self-interference and inter-BS-interference channel matrices, respectively; $\zeta_m>0$ and $\eta_{l,m}>0$ indicate the residual self-interference and inter-BS-interference suppression levels, respectively; $\mathbf{n}_{r,m}[n] \sim \mathcal{CN}(\mathbf{0}, \sigma_r^2 \mathbf{I})$ is the sensing AWGN vector for BS $m$.

Note that we adopt the normalized co-site leakage model $\mathbf{H}^{\mathrm{SI}}_m[n]=\mathbf{I}_S$ for the SI channel \cite{ref06}. For $l\neq m$, the IBI channel is modeled as $\mathbf{H}^{\mathrm{IBI}}_{l,m}[n]=d_{l,m}^{-1}\mathbf{G}_{l,m}$, where $d_{l,m}=\|\mathbf{v}_l-\mathbf{v}_m\|$ and the entries of $\mathbf{G}_{l,m}$ are independently drawn from $\mathcal{CN}(0,1)$. Each $\mathbf{G}_{l,m}$ realization is generated at the beginning of an episode and remains fixed within that episode; the residual SI and IBI powers are subsequently controlled by $\zeta_m$ and $\eta_{l,m}$ in \eqref{eq:received_sensing_signal}, respectively.

For the received signal vector $\mathbf{y}_{r,m}[n]$, BS $m$ employs a receive beamforming $\mathbf{u}_{m,j}[n] \in \mathbb{C}^{S \times 1}$ to extract the echo of S-UAV $j\in{\cal K}_s$. Therefore, at each time slot $n\in{\cal N}$, the sensing SINR of S-UAV $j\in{\cal K}_s$ at BS $m$ is given by 
\begin{equation}\label{eq:sinr_rayleigh}
\gamma_{m,j}^s[n] = \frac{\mathbf{u}_{m,j}^H[n] \, \mathbf{E}_{m,j}[n] \, \mathbf{u}_{m,j}[n]}{\mathbf{u}_{m,j}^H[n] \, \mathbf{F}_{m,j}[n] \, \mathbf{u}_{m,j}[n]},
\end{equation}
where we have
\begin{align*}
\mathbf{E}_{m,j}[n] & \triangleq \sigma_t^2 \mathbf{A}^s_{m,j}[n] \mathbf{X}_m[n]\mathbf{A}_{m,j}^{sH}[n] \\ %\label{eq:E_matrix} \\
\mathbf{F}_{m,j}[n] &\triangleq \zeta_m \mathbf{H}^\text{SI}_m[n] \mathbf{X}_m[n] \mathbf{H}^{\text{SI}H}_m[n] \notag \\
&+ \sum_{l \in \mathcal{M} \setminus \{m\}} \eta_{l,m} \mathbf{H}^{\text{IBI}}_{l,m}[n] \mathbf{X}_l[n]  \mathbf{H}^{\text{IBI}H}_{l,m}[n] \notag \\
& + \sum_{i \in \mathcal{K}_c} \sigma_t^2\mathbf{A}^c_{m,k}[n] \mathbf{X}_m[n] \mathbf{A}^{cH}_{m,k}[n] \notag \\ 
& + \sum_{k\in \mathcal{K}_s, k \neq j} \sigma_t^2\mathbf{A}^s_{m,k}[n] \mathbf{X}_m[n] \mathbf{A}_{m,k}^{sH}[n] + \sigma_r^2 \mathbf{I}_S  \\ %\label{eq:F_matrix}\\
\mathbf{X}_m[n] &= 
\sum_{i \in \mathcal{K}_c} \mathbf{w}^c_{m,i}[n]\mathbf{w}^{cH}_{m,i}[n] + \sum_{j \in \mathcal{K}_s} \mathbf{w}^s_{m,j}[n]\mathbf{w}^{sH}_{m,j}[n].
\end{align*}
Note that $\mathbf{E}_{m,j}[n]$ captures the power characteristics of the desired echo signal, and $\mathbf{F}_{m,j}[n]$ synthesizes the residual self-interference, inter-BS cross-interference, and receiver noise. 

We consider distributed data-level non-coherent fusion, where BS $m$ obtains the local sensing statistic $z_{m,j}[n]=\mathbf{u}_{m,j}^H[n]\mathbf{y}_{r,m}[n]$ for S-UAV $j$, and normalizes it by the local interference-plus-noise standard deviation before reporting it to the central controller. The local sensing SINR in \eqref{eq:sinr_rayleigh} is the ratio of the desired echo power to the residual interference-plus-noise power. Since the local statistics are normalized to unit local interference-plus-noise power, the sum of the local sensing SINRs serves as a tractable surrogate for the fused output SINR under non-coherent combining, i.e., $\gamma_j^s[n]\triangleq\sum_{m\in\mathcal{M}}\gamma_{m,j}^s[n]$. Accordingly, the sensing SINR constraints are imposed on the aggregated metric as
\begin{equation}
\gamma_j^s[n] \geq \Gamma_j,~\forall j\in{\cal K}_s,
\label{eq:sensing_sinr}
\end{equation}
where $\Gamma_j>0$ is the prescribed sensing quality of service (QoS) threshold for tracking S-UAV $j\in{\cal K}_s$. We assume that the residual cross-covariances between the BSs,  conditioned on the current geometry and transmit beamformers, are negligible, owing to their geographic separation and the directional receive beamforming.

\section{Problem Formulation}
\label{sec:problem_formulation}
In this work, we aim to maximize the downlink average sum rate of C-UAVs by jointly optimizing the BS ISAC transmit/receive beamformers and the UAV trajectories, subject to the per-BS transmit power budgets, the sensing SINR constraints, and the strigent UAV kinematic constraints. To this end, the multi-BS ISAC joint optimization problem, denoted by (P0), is formulated as:
\begin{subequations} \label{eq:optimization_problem}
\begin{align} 
\text{(P0):} \quad & \max_{\substack{\mathcal{W}^c, \mathcal{W}^s, \mathcal{U}, \mathcal{Q}}} \; \frac{1}{N} \sum_{n=0}^{N-1} R[n] \label{eq:opt_obj} \\
\text{s.t.} \quad & \gamma_j^s[n] \geq \Gamma_j, \, \forall j \in \mathcal{K}_s, n \in \mathcal{N}, \label{eq:opt_sensing} \\
& \eqref{eq:trajectory1},\eqref{eq:trajectory2}, \eqref{eq:trajectory3}  ~ \text{and}~ \eqref{eq:ISAC_power},
\end{align}
\end{subequations}
where the sets $\mathcal{W}^c \triangleq \{\mathbf{w}^c_{m,i}[n], \forall m \in \mathcal{M}, i \in \mathcal{K}_c, n \in \mathcal{N}\}$ and $\mathcal{W}^s \triangleq \{\mathbf{w}^s_{m,j}[n], \forall m \in \mathcal{M}, j \in \mathcal{K}_s, n \in \mathcal{N}\}$ collect all the BS transmit beamforming vectors for downlink communication and sensing, respectively; $\mathcal{U} \triangleq \{\mathbf{u}_{m,j}[n], \forall m \in \mathcal{M}, j \in \mathcal{K}_s, n \in \mathcal{N}\}$ collects the BS receive beamforming vectors; and $\mathcal{Q} \triangleq \{\mathbf{q}^c_i[n], \mathbf{q}^s_j[n], \forall i \in \mathcal{K}_c, j\in \mathcal{K}_s, n \in \mathcal{N}\}$ denotes the spatial trajectory vectors of the C-UAVs/S-UAVs.

\subsection{Closed-Form Solution for Receive Beamformers $\{\mathbf{u}^*_{m,j}[n]\}$}\label{subsubsec:closed_form}

Let $\lambda_{m,j}^{\max}[n]$ and $\mathbf{v}_{m,j}^{\max}[n]$ denote the maximum eigenvalue and its corresponding principal eigenvector of $\mathbf{M}_{m,j}[n]\triangleq \mathbf{F}_{m,j}^{-1/2}[n]\mathbf{E}_{m,j}[n] \mathbf{F}_{m,j}^{-1/2}[n]$, respectively. Based on the generalized Rayleigh quotient theorem \cite{new20}, we establish the following proposition. 

\begin{proposition} {[Optimal Receive beamformers $\{\mathbf{u}^*_{m,j}[n]\}$]}
\label{prop:optimal_filter}
Under the local sensing-SINR definition in \eqref{eq:sinr_rayleigh} and the aggregated sensing-SINR metric, the optimal receive beamformer for sensing S-UAV $j\in\mathcal{K}_s$ at BS $m$ and slot $n$ is
\begin{align}
\mathbf{u}_{m,j}^*[n] =\mathbf{F}_{m,j}^{-1/2}[n] \, \mathbf{v}_{m,j}^{\max}[n],
\label{eq:optimal_filter}
\end{align}
and the maximum local sensing SINR is $\lambda_{m,j}^{\max}[n]=\max_{\mathbf{u}_{m,j}[n] \neq \mathbf{0}} \gamma_{m,j}^s[n]$.
\end{proposition}

\begin{IEEEproof}
By substituting $\tilde{\mathbf{u}}_{m,j}[n] = \mathbf{F}_{m,j}^{1/2}[n] \, \mathbf{u}_{m,j}[n]$ to \eqref{eq:sinr_rayleigh}, the Rayleigh quotient is transformed into a standard form as
\begin{equation}
\gamma_{m,j}^s[n] = \frac{\tilde{\mathbf{u}}_{m,j}^H[n] \, \mathbf{M}_{m,j}[n] \, \tilde{\mathbf{u}}_{m,j}[n]}{\tilde{\mathbf{u}}_{m,j}^H[n] \, \tilde{\mathbf{u}}_{m,j}[n]}.
\label{eq:standard_rayleigh}
\end{equation}
 Based on the generalized Rayleigh quotient theorem \cite{new20}, the maximum value of a standard Rayleigh quotient equals the largest eigenvalue of the matrix $\mathbf{M}_{m,j}[n]$, which is achieved when $\tilde{\mathbf{u}}_{m,j}[n]$ aligns with the corresponding principal eigenvector $\mathbf{v}_{m,j}^{\max}[n]$. Restoring the variable substitution yields \eqref{eq:optimal_filter}. Since the Rayleigh quotient is invariant to any non-zero complex scalar scaling of the receive beamforming, the normalization constant does not affect the SINR value.
\end{IEEEproof}

\subsection{Problem Reformulation}
By substituting the optimal receive beamformers $\{\mathbf{u}_{m,j}^*[n]\}$ in Proposition~\ref{prop:optimal_filter} into (P0), we obtain the equivalent problem of joint multi-BS ISAC beamforming and trajectory design, denoted by (P1), as
\begin{subequations} \label{eq:p1}
\begin{align}
\text{(P1):}  & \max_{\mathcal{W}^c, \mathcal{W}^s, \mathcal{Q}}  \frac{1}{N} \sum_{n=0}^{N-1} R[n] \label{eq:p1_obj} \\
\text{s.t.} \quad & \sum_{m \in \mathcal{M}} \lambda_{m,j}^{\max}[n] \geq \Gamma_j, \, \forall j \in \mathcal{K}_s, n \in \mathcal{N}, \label{eq:p1_sensing} \\
& \eqref{eq:trajectory1},\eqref{eq:trajectory2}, \eqref{eq:trajectory3}  ~ \text{and}~ \eqref{eq:ISAC_power}.
\end{align}
\end{subequations}
Note that, due to the multiple quadratic fractional terms associated with the beamforming and trajectory variables, (P1) is nonconvex and is thus challenging to solve. Moreover, the time-varying channels $\{\mathbf{h}^c_{m,i}[n],\mathbf{h}^s_{m,j}[n]\}$ and the tight coupling between the continuous UAV trajectories and the ISAC transmit beamforming vectors further increase the dimensionality of the problem, rendering conventional optimization methods intractable for low-latency online execution. DRL offers an attractive alternative by shifting the computational burden to offline training, enabling millisecond-level online action inference while inherently accommodating the time-varying channels and the sequential decision coupling. Nevertheless, the heterogeneous nature of the beamforming and trajectory actions makes a monolithic single-agent policy difficult to train efficiently. Motivated by these observations, we next leverage the MADRL technique to develop a structure-aware dual-agent framework for solving (P1) (or equivalently, (P0)).

%==============================================================================
\begin{figure*}
  \centering
  \includegraphics[width=0.85\textwidth]{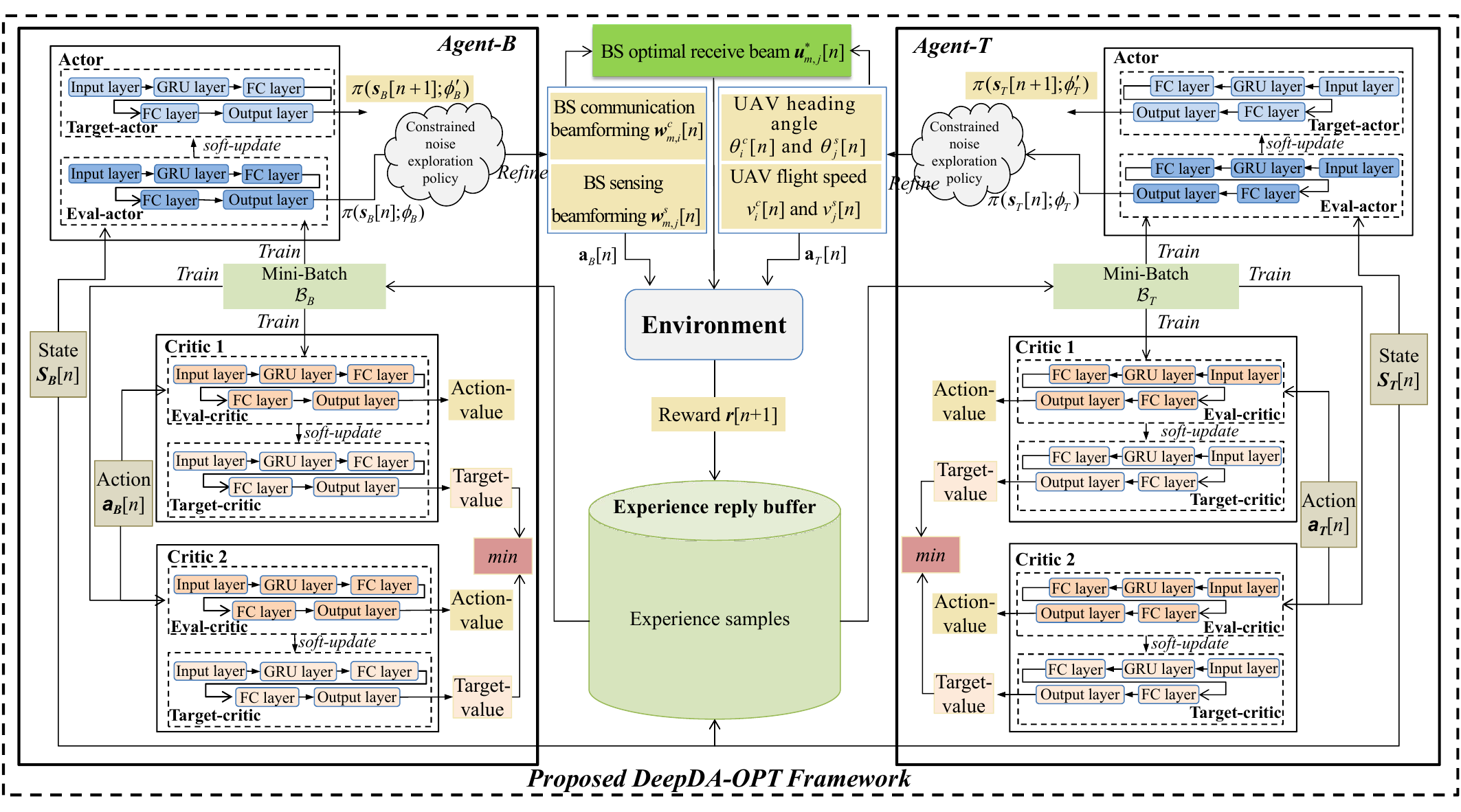}
  \caption{Proposed DeepDA-OPT framework for multi-BS ISAC beamforming and trajectory design.}
  \label{fig:framework}
\end{figure*}
%==============================================================================

\section{Dual-Agent MDP Formulation}
\label{sec:mdp_reformulation}
To facilitate the MADRL solution, in this section we recast (P1) as a dual-agent Markov decision process (MDP), consisting of a beamforming agent (Agent-B) and a trajectory agent (Agent-T). Specifically, Agent-B is responsible for adapting the ISAC transmit beamforming vectors to the highly dynamic communication and sensing CSIs in each time slot, whereas Agent-T is responsible for managing the flight kinematics and spatial topologies of all the C-UAVs/S-UAVs to fulfill the flight-missions and collision-avoidance requirements.

\subsection{Formulation of Agent-B}
\subsubsection{State Space $\mathcal{S}_B$} 
The state space $\mathcal{S}_B$ of Agent-B is a set of all possible channel states. At each time slot $n\in{\cal N}$, the state $\mathbf{s}_B[n]\in{\cal S}_B$ observed by Agent-B consists of the downlink channels from the BSs to all the UAVs, the inter-BS interference channels, and the self-interference channels at the BSs. Specifically, $\mathbf{s}_B[n]$ is given by
\begin{equation}
\mathbf{s}_B[n] = (\{ \mathbf{h}^c_{m,i}[n]\},\{\mathbf{h}^s_{m,j}[n]\},\{\mathbf{H}^{\text{IBI}}_{l,m}[n]\}, \{\mathbf{H}^{\text{SI}}_m[n]\}).
\label{eq:state_beamforming}
\end{equation}

\begin{remark}
 While the instantaneous physical CSI observation available to Agent-B is specified in~\eqref{eq:state_beamforming}, both the actor and critic adopt recurrent network architectures, so the information they effectively exploit at time slot $n$ also includes the GRU hidden state $\mathbf{h}^{\mathrm{GRU}}_B[n]$, which recursively encodes the sequence of past observations and is reset at the beginning of each episode. Hence, both the actor and the critic effectively condition on the current CSI together with its recurrent temporal context, rather than the instantaneous CSI alone. For notational simplicity, the dependence on the GRU hidden state is suppressed in the subsequent actor and critic expressions unless otherwise specified.
\end{remark}

\subsubsection{Action Space $\mathcal{A}_B$} 
The action space $\mathcal{A}_B$ of Agent-B consists of the ISAC transmit beamforming vectors applied at all the $M$ BSs. At each slot $n\in{\cal N}$, the action $\mathbf{a}_B[n]\in{\cal A}_B$ executed by Agent-B comprises the communication beamforming vectors for C-UAVs and the sensing beamforming vectors for S-UAVs. Specifically, $\mathbf{a}_B[n]$ is given by
\begin{equation}
\mathbf{a}_B[n] = ( \{\mathbf{w}^c_{m,i}[n]\}, \{\mathbf{w}^s_{m,j}[n]\}).
\label{eq:action_beamforming}
\end{equation}

\subsection{Formulation of Agent-T}
Agent-T is responsible for managing the macroscopic flight kinematics and spatial topologies of all the C-UAVs and S-UAVs to satisfy the flight missions and collision avoidance requirements.

\subsubsection{State Space $\mathcal{S}_T$} 
The state space $\mathcal{S}_T$ of Agent-T consists of the spatial and temporal information required for trajectory planning. At each time slot $n$, the state $\mathbf{s}_T[n]\in{\cal S}_T$ observed by Agent-T comprises the current and final locations of all the C-UAVs and S-UAVs, as well as the their temporal urgency factors. Specifically, $\mathbf{s}_T[n]$ is given by
\begin{equation}
\mathbf{s}_T[n] = ( \{ \mathbf{q}^c_i[n]  , \mathbf{{q}}_i^c[N],\delta^c_i[n]\},\{ \mathbf{q}^s_j[n], \mathbf{{q}}_j^s[N],\delta^s_j[n]\} ),
\label{eq:state_trajectory}
\end{equation}
where $\delta^c_i[n]$ and $\delta^s_j[n]$ denotes the temporal urgency factors of C-UAV $i$ and S-UAV $j$, defined as the ratios of the remaining straight-line distances to the maximum distances that can be covered within the remaining mission time, i.e., $\delta^c_i[n] \triangleq \frac{\|\mathbf{q}^c_i[n] - \mathbf{{q}}^c_i[N]\|}{V_{\max}(N - n)\Delta t}$ and $\delta^s_j[n] \triangleq \frac{\|\mathbf{q}^s_j[n] - \mathbf{{q}}_j^s[N]\|}{V_{\max}(N - n)\Delta t}$, respectively. In particular, $\delta^c_i[n]>1$ indicates that C-UAV $i$ cannot reach its destination within the remaining mission time.

\subsubsection{Action Space $\mathcal{A}_T$} 
The action space $\mathcal{A}_T$ of Agent-T consists of the continuous flight control variables of all the C-UAVs and S-UAVs. At time slot $n$, the action $\mathbf{a}_T[n]\in{\cal A}_T$ determines the heading angle $\theta^c_i[n] \in [0, 2\pi)$ and flight speed $v^c_i[n] \in [0, V_{\max}]$ for each C-UAV $i\in{\cal K}_c$, as well as the heading angle $\theta^s_i[n] \in [0, 2\pi)$ and flight speed $v^s_i[n] \in [0, V_{\max}]$ for each S-UAV $j\in{\cal K}_s$. Thus, $\mathbf{a}_T[n]$ is given by
\begin{equation}
\mathbf{a}_T[n] = (\{\theta^c_i[n], v^c_i[n]\},\{\theta^s_j[n], v^s_j[n]\}).
\label{eq:action_trajectory}
\end{equation}
Since the exploration noise may drive the actions outside the feasible sets, both $\mathbf{a}_B[n]$ and $\mathbf{a}_T[n]$ are rectified on the environment side by a constraint-aware exploration policy, as detailed in Section \ref{subsec:datd3_architecture}).

\subsection{State Transition Probabilities}
We define the joint state and action as $\mathbf{s}[n]\triangleq(\mathbf{s}_B[n],\mathbf{s}_T[n])$ and $\mathbf{a}[n]\triangleq(\mathbf{a}_B[n],\mathbf{a}_T[n])$, respectively, with the joint state and action spaces given by ${\cal S}={\cal S}_B\times{\cal S}_T$ and ${\cal A}={\cal A}_B\times{\cal A}_T$. The joint transition state evolves according to the transition law $\mathcal{P}(\mathbf{s}[n+1] \mid \mathbf{s}[n],\mathbf{a}[n])$. In particular, the heading and speed actions of Agent-T update the positions of all UAVs, which in turn determine the geometry-dependent CSI observed by Agent-B. The UAV-position update is given by
\begin{subequations}\label{eq:transition_kinematic_update}
\begin{align}
\mathbf{q}_i^c[n+1] &= \mathbf{q}_i^c[n] + v_i^c[n]\Delta t \tilde{\boldsymbol{\theta}}^c_i[n], \forall i \in \mathcal{K}_c, \label{eq:transition_kinematic_c} \\
\mathbf{q}_j^s[n+1] &= \mathbf{q}_j^s[n] + v_j^s[n]\Delta t \tilde{\boldsymbol{\theta}}^s_j[n], \forall j \in \mathcal{K}_s. \label{eq:transition_kinematic_s}
\end{align}
\end{subequations}
where $\tilde{\boldsymbol{\theta}}^c_i[n]=[\cos\theta_i^c[n],\sin\theta_i^c[n]]^T$ and $\tilde{\boldsymbol{\theta}}^s_j[n]=[\cos\theta_j^s[n],\sin\theta_j^s[n]]^T$.

By the chain rule, the joint transition law can be factorized as
\begin{align}\label{eq:joint_state_transition}
&\mathcal{P}(\mathbf{s}[n+1]\mid\mathbf{s}[n],\mathbf{a}[n]) = \mathcal{P}_T(\mathbf{s}_T[n+1]\mid\mathbf{s}_T[n],\mathbf{a}_T[n]) \notag \\
&~~~~~ \times\mathcal{P}_B(\mathbf{s}_B[n+1]\mid \mathbf{s}_T[n+1]),
\end{align}
where ${\cal P}_T$ and ${\cal P}_B$ denote the trajectory-state and channel-state transition factors, respectively. As shown in \eqref{eq:joint_state_transition}, the channel-state transition depends only on $(\mathbf{s}_T[n],\mathbf{a}_T[n])$, since the UAV kinematics are not affected by the channel state and the beamforming action; the channel-state transition depends only on the next trajectory state $\mathbf{s}_T[n+1]$, since the UAV locations determine the next-slot geometric channels in the adopted model. Consequently, the joint transition law is independent of the beamforming action $\mathbf{a}_B[n]$, which affects only the instantaneous ISAC performance and hence the rewards of both agents.

\subsection{Reward Function}
To align with the objective of (P1), the reward is designed to maximize the downlink sum rate of the C-UAVs while penalizing the violations of the collision-avoidance, sensing, and flight-urgency constraints. Specifically, at each time slot $n\in{\cal N}$, the immediate reward $r[n]$ shared by Agent-B and Agent-T is given by 
\begin{equation}
r[n] = \omega_1 R[n] -\omega_2 P_{col}[n] - \omega_3 P_{sen}[n] -\omega_4 P_{urg}[n],
\label{eq:reward}
\end{equation}
where $\omega_k>0$, $k=1,2,3,4$, are the reward weighting factors. Here, $P_{col}[n] \triangleq \sum_{i\neq i'\in{\cal K}_c} \mathbbm{1}(d^c_{i,i'} < d_{\min})+\sum_{j\neq j'\in{\cal K}_s} \mathbbm{1}(d^s_{j,j'} < d_{\min})+\sum_{i\in{\cal K}_c,j\in{\cal K}_s} \mathbbm{1}(d_{i,j} < d_{\min})$ is the collision penalty, with $d^c_{i,i'}=\|\mathbf{q}^c_i[n] - \mathbf{q}^c_{i'}[n]\|$, $d^s_{j,j'}=\|\mathbf{q}^s_j[n] - \mathbf{q}^s_{j'}[n]\|$, $d_{i,j}=\|\mathbf{q}^c_i[n] - \mathbf{q}^s_j[n]\|$ denoting the pairwise C-UAV, S-UAV, and C-UAV--S-UAV distances, respectively; $P_{sen}[n]\triangleq \sum_{j \in \mathcal{K}_s}
(\Gamma_j - \sum_{m \in \mathcal{M}} \lambda_{m,j}^{\max}[n])^+$ is the sensing-SINR violation penalty, where $(x)^+=\max(x,0)$; $P_{urg}[n] \triangleq \sum_{i\in \mathcal{K}_c} (\delta^c_i[n] - \bar{\delta}^c)^+ + \sum_{j\in \mathcal{K}_s} (\delta^s_j[n] - \bar{\delta}^s)^+$ is the temporal-urgency penalty, where $\bar{\delta}^c<1$ and $\bar{\delta}^s<1$ are the corresponding urgency thresholds of C-UAV and S-UAV, respectively, chosen below one to provide an early-warning margin before the destinations become strictly unreachable.

\begin{remark}
Based on the transition factorization in \eqref{eq:transition_kinematic_update}, the trajectory process is independent of the beamforming action. Hence, $P_{col}[n]$ and $P_{urg}[n]$, which depend only on the UAV trajectories, provides no policy-gradient signal to Agent-B. In contrast, $R[n]$ and $P_{sen}[n]$ depend jointly on the channel states (which are determined by the UAV geometry) and the beamforming action, thereby coupling the trajectory and beamforming decisions through the shared rewards. The shared reward thus aligns both agents with the system-level objective of (P1) without requiring every reward component to depend on the actions of both agents.
\end{remark}

\section{Proposed DeepDA-OPT Scheme}
\label{sec:proposed_scheme}
 To solve the formulated dual-agent MDP (or equivalent,(P1)), we propose the DeepDA-OPT Scheme as illustrated in Fig.~\ref{fig:framework}. 

\subsection{DeepDA-OPT Architecture}
\label{subsec:datd3_architecture}

 Unlike conventional DDPG algorithms, which may suffer from action-value overestimation \cite{ref23}, the TD3 algorithm is adopted in the proposed DeepDA-OPT scheme. Moreover, the actor and critic networks of both Agent-B and Agent-T adopt a recurrent neural network (RNN) architecture to capture the temporal correlations of the high-mobility C-UAV/S-UAV channels.

\subsubsection{Actor Network and Twin Critic Networks}
 In the proposed DeepDA-OPT scheme, the actor-critic architecture of each agent consists of one evaluation actor network and a pair of two evaluation twin critic networks. For Agent-B, denote by $(\pi(\phi_B),Q(\psi_{B1}),Q(\psi_{B2}))$ and $(\pi(\phi'_B),Q(\psi'_{B1}),Q(\psi'_{B2}))$ the evaluation and target actor-critic networks, respectively, where $(\phi_B, \psi_{B1},\psi_{B2},\phi'_B, \psi'_{B1},\psi'_{B2})$ are learnable network parameters. The networks of Agent-T are denoted similarly, with parameters $(\phi_T, \psi_{T1},\psi_{T2},\phi'_T, \psi'_{T1},\psi'_{T2})$. For both agents, the target actor-critic networks are utilized to compute stable target $Q$-values during the training of the evaluation critics, thereby stabilizing the training process. Moreover, each actor and critic network consists of an input layer, a gated recurrent unit (GRU) layer, two fully connected (FC) layers, and an output layer, where the GRU layer is incorporated to capture the cross-slot temporal correlation inherent in the highly dynamic UAV environment \cite{new11}.

\subsubsection{Action Output and Physical Mapping}
Denote by $\tilde{\mathbf{a}}_B[n]\in[-1,1]^{2MKS}$ and $\tilde{\mathbf{a}}_T[n]\in[-1,1]^{2K}$ the normalized action vectors of Agent-B and Agent-T, respectively, where $K=K_c+K_x$. They are obtained element-wise clipping the policy outputs plus the exploration noise to $[-1,1]$, i.e.,
\begin{subequations}
\begin{align}
    &\tilde{\mathbf{a}}_B[n] = \text{clip}(\pi(\mathbf{s}_B[n];\phi_B) + \boldsymbol{\epsilon}_B[n], -1, 1)  \\
    &\tilde{\mathbf{a}}_T[n] = \text{clip}(\pi(\mathbf{s}_T[n];\phi_T) + \boldsymbol{\epsilon}_T[n], -1, 1)
\end{align}
\end{subequations}
where $\pi(\cdot;\phi_B)$ and $\pi(\cdot;\phi_T)$ denotes the deterministic policies parameterized by $\phi_B$ and $\phi_T$, respectively, and $\boldsymbol{\epsilon}_B[n]$ and $\boldsymbol{\epsilon}_T[n]$ are the exploration noises.

For Agent-B, the first $2MK_cS$ elements of $\tilde{\mathbf{a}}_B[n]$, arranged BS-wise, are grouped in pair, with each adjacent pair represented the real and imaginary parts of one complex entry, to construct the un-normalized downlink communication beamforming vectors $\{\tilde{\mathbf{w}}^c_{m,i}[n]\}_{m\in{\cal M},i\in{\cal K}_c}$ for the C-UAVs; the remaining $2MK_sS$ elements are grouped likewise to construct the un-normalized sensing beamforming vectors $\{\tilde{\mathbf{w}}^s_{m,j}[n]\}_{m\in{\cal M},j\in{\cal K}_s}$ of the S-UAVs. Then, the physical beamforming vectors are obtained by per-BS power normalization, i.e.,
 \begin{subequations}\label{eq.power_normalization}
 \begin{align}
&\mathbf{w}_{m,i}^c[n] = \tilde{\sigma}_m[n] \tilde{\mathbf{w}}_{m,i}^c[n] 
\\ 
&\mathbf{w}_{m,j}^s[n] = \tilde{\sigma}_m[n] \tilde{\mathbf{w}}_{m,j}^s[n],
\label{eq:power_normalization}
\end{align}
\end{subequations}
where $\tilde{\sigma}_m[n] = \sqrt{{P_{\max}}/{\tilde{P}[n]}}$ is the power scaling factor, and $P[n]=\sum_{i \in \mathcal{K}_c} \|\tilde{\mathbf{w}}_{m,i}^c[n]\|^2 + \sum_{j \in \mathcal{K}_s} \|\tilde{\mathbf{w}}_{m,j}^s[n]\|^2$ denotes the un-normalized transmit power of BS $m$. It is observed that the normalization in \eqref{eq.power_normalization} satisfies the per-BS power constraint in (P1) with equality.

For Agent-T, the clipped logical action vector $\tilde{\mathbf{a}}_T[n]\in[-1,1]^{2K}$ is decomposed into $K$ (heading, speed) pairs. The first $2K_c$ elements, grouped in adjacent pairs, yield $\{(\tilde{\theta}_i^c[n], \tilde{v}_i^c[n])\}_{i\in{\cal K}_c}$ for the C-UAVs, and the remaining $2K_s$ elements yield $\{(\tilde{\theta}_j^s[n], \tilde{v}_j^s[n])\}_{j\in{cal K}_s}$ for the S-UAVs. Define by $N_{i}^c[n] = \lceil \|\mathbf{q}_i^c[n] - \mathbf{q}_i^c[N]\|/(V_{\max} \Delta t)\rceil$ and $N_{j}^s[n] = \lceil \|\mathbf{q}_j^s[n] - \mathbf{q}_j^s[N]\|/(V_{\max} \Delta t) \rceil$ the minimum numbers of time slots required for C-UAV $i$ and S-UAV $j$ to reach their destinations via straight-line flight at $V_{\max}$, respectively. For each C-UAV $i\in{\cal K}_c$, if $N_{i}^c[n]<N-n-1$, then the logical pair is mapped to the physical action as $\theta_i^c[n]=(\tilde{\theta}_i^c[n] +1)\pi$ and $v_i^c[n]=(\tilde{v}_i^c[n] + 1)V_{\max}/2$; otherwise, the action is overridden by the destination-oriented flight $\theta_i^c[n]=\bar{\theta}_i^{c}[n]$ and $v_j^c[n]=\min\{V_{\max},\|\mathbf{q}_i^c[n] - \mathbf{q}_i^c[N]\|/\Delta t\}$, where $\bar{\theta}_i^{c}[n] \triangleq \mathrm{arctan2}(y_i^c[N] - y_i^c[n], {x}_i^c[N] - x_i^c[n])$ is the heading angle of the straight line toward the destination. The same mapping and override procedures apply to each S-UAV $j\in{\cal K}_s$, with $\bar{\theta}_j^{s}[n]$ defined analogously. Hence, Agent-T can freely explore trajectory designs to improve the ISAC geometry when the time budget is sufficient, and the learned actions are overridden by the destination-oriented flight otherwise.

\begin{remark}
The switching condition $N_i^c[n]<N-n-1$ is designed to balance geometric exploration against guaranteed mission completion. When the remaining time budget is sufficient, Agent-T may exploit the redundancy to improve the ISAC geometry; conversely, excessive maneuvering away from the destinations consumes the remaining margin. The one-slot margin in the switching condition (and analogously for the S-UAVs) guarantees terminal reachability: even if a UAV moves away from its destination by up to one slot of flight at $V_{\max}$, the the remaining horizon still suffices for the subsequent straight-line flight. As an example, under the default setting with a straight-line distance of 300~m to each destination, $V_{\max}\Delta t=20$~m, and $N=20$, 15 full-speed slots are required, leaving five slots of redundancy, of which one slot is reserved as the reachability margin and about four slots remain for free trajectory maneuvering. Once this margin can no longer be preserved, the learned action is overridden by the destination-oriented flight.
\end{remark}

\subsection{Online Experience Interaction Phase}
 We adopt a hierarchical experience replay mechanism during the online interaction phase of DeepDA-OPT. Denote by $\mathcal{D}$ the experience reply buffer, which stores the experience samples generated during training, where each experience sample corresponds to one complete episode. Specifically, each experience sample $\mathcal{E}\in\mathcal{D}$ collects all the experience tuples of all time slots and is given by
\begin{align}
    \mathcal{E} = \{\mathcal{T}[0],\mathcal{T}[1],...,\mathcal{T}[N-1]\},
\end{align} 
where $\mathcal{T}[n] \triangleq (\mathbf{s}_B[n],\mathbf{s}_T[n],\mathbf{a}_B[n],\mathbf{a}_T[n], r[n], \mathbf{s}_B[n+1],\mathbf{s}_T[n+1], d[n])$ is the experience tuple generated at each time slot $n$ and $d[n]$ is the episode termination flag with $d[n]=0$ for $n\in{\cal N}\setminus\{N-1\}$ and $\delta[N-1]=1$.

\subsection{Offline Training Phase}
 During the training phase, the two agent-specific mini-batches are obtained by extracting the agent-specific state-action components from the same sampled episode $\mathcal{E}$; hence, the two mini-batches share the same time indices, the same reward $r[n]$, ad the same termination flag $d[n]$ at each time slot. DeepDA-OPT deviates from the standard CTDE architecture in the sense that each critic takes only its own agent's state-action pair as input, rather than the joint state and joint action. The shared reward and the episode-aligned joint experience coordinate the two agents in the sense that both are optimized toward the same system-level objective of (P1), while the clipped double Q-learning, delayed policy updates, and soft target updates of TD3 mitigate, rather than eliminate, the non-stationarity arising from the simultaneous policy updates of the two agents.

\subsubsection{Critic Network Update via Clipped Double Q-Learning}
Let $\pi_B\triangleq \pi(\cdot;\phi_B)$ and $\pi_T\triangleq\pi(\cdot;\phi_T)$ denote the policies of Agent-B and Agent-T, respectively. We define the recurrent information available to the corresponding GRU-based critics as $\boldsymbol{\chi}_B[n]\triangleq(\mathbf{s}_B[n],\mathbf{h}^{\mathrm{GRU}}_B[n])$ and $\boldsymbol{\chi}_T[n]\triangleq(\mathbf{s}_T[n],\mathbf{h}^{\mathrm{GRU}}_T[n])$, where the hidden states are updated recurrently along each episode. The finite-horizon action-value functions are conditioned on the recurrent information and current action, i.e.,
\begin{subequations}\label{eq:q_function}
\begin{align}
Q_{B,n}^{\pi_B}\left(\boldsymbol{\chi}_B[n],\mathbf{a}_B[n]\right) &= \mathbb{E}_{\pi_B}\left[G[n] \mid \boldsymbol{\chi}_B[n],\mathbf{a}_B[n]\right] \label{eq:q_function_B} \\
Q_{T,n}^{\pi_T}\left(\boldsymbol{\chi}_T[n],\mathbf{a}_T[n]\right) &= \mathbb{E}_{\pi_T}\left[G[n] \mid \boldsymbol{\chi}_T[n],\mathbf{a}_T[n]\right], \label{eq:q_function_T}
\end{align}
\end{subequations}
where $G[n] \triangleq \sum_{t=0}^{N-1-n} \gamma^{t} r[n+t]$ denotes the finite-horizon cumulative return, $\gamma \in (0, 1)$ is the discount factor, and the expectation are taken under the corresponding policy and the system dynamics. The subscript $n$ makes the finite-horizon stage dependence explicit. Accordingly, the one-step finite-horizon recursions are respectively given by 
\begin{subequations}\label{eq:bellman_equation}
\begin{align}
&Q_{B,n}^{\pi_B} = \mathbb{E}_{\pi_B}[r[n]+\gamma Q_{B,n+1}^{\pi_B}\mid \boldsymbol{\chi}_B[n],\mathbf{a}_B[n]] \label{eq:bellman_B} \\
& Q_{T,n}^{\pi_T} = \mathbb{E}_{\pi_T}[r[n]+\gamma Q_{T,n+1}^{\pi_T}\mid \boldsymbol{\chi}_T[n],\mathbf{a}_T[n]].\label{eq:bellman_T}
\end{align}
\end{subequations}
 At the terminal slot $n=N-1$, no further value is bootstrapped, which is consistent with the termination flag $d[N-1]=1$. In the subsequent update rules, $Q_B(n;\psi_B)$ and $Q_T(n;\psi_T)$ denotes the recurrent critics parameterized by $\psi_B$ and $\psi_T$, evaluated at the slot-$n$ recurrent information and action.

 For each mini-batch ${\cal B}$ sampled from the replay buffer $\cal D$, denote by $y_B^{(k)}[n]$ and $y_T^{(k)}[n]$ the temporal difference (TD) targets for the evaluation critics of Agent-B and Agent-T, respectively, where $k=1,2,...,|{\cal B}|$. Let $\phi'_B$ and $\phi'_T$ denote the target actor parameters, which generate the next-slot actions $\tilde{\mathbf{a}}_B[n+1]=\text{clip}(\pi(\mathbf{s}_B[n+1;\phi'_B],-1,1))$ and $\tilde{\mathbf{a}}_T[n+1]=\text{clip}(\pi(\mathbf{s}_T[n+1;\phi'_B],-1,1))$, and let $\psi'$ the target critic parameters. Based on the clipped double Q-learning principle, the TD targets are given by $y_B^{(k)}[n] = r^{(k)}[n] + \gamma (1-d^{(k)}[n])\min ( Q(n+1;{\psi'_{B1}}),Q(n+1;{\psi'_{B2}}))$ and $y_T^{(k)}[n] = r^{(k)}[n] + \gamma  (1-d^{(k)}[n]) \min ( Q(n+1;{\psi'_{T1}}),Q(n+1;{\psi'_{T2}}))$ for $n\in{\cal N}$, where $Q(n+1;\psi')$ is evaluated at the slot-($n+1$) recurrent information and the target-actor action; at the terminal slot, $d^{(k)}[n-1]=1$ and the bootstrap term vanishes.

 To evaluate the network approximations, the loss functions of the twin evaluation critics are given by 
\begin{subequations}\label{eq:critic_loss}
\begin{align}
\mathcal{L}_{B,\ell} &= \frac{1}{|\cal B|} \sum_{k=1}^{|\cal B|} \sum_{n=0}^{N-1} ( y_B^{(k)}[n] - Q(n;\psi_{B,\ell}))^2, \ell=1,2 \label{eq:critic_loss_B} \\
\mathcal{L}_{T,\ell} &= \frac{1}{|\cal B|} \sum_{k=1}^{|\cal B|} \sum_{n=0}^{N-1} ( y_T^{(k)}[n] - Q(n; \psi_{T,\ell}) )^2, \ell=1,2. \label{eq:critic_loss_T}
\end{align}
\end{subequations}

To minimize the loss in \eqref{eq:critic_loss}, the parameters $(\psi_{B,1},\psi_{B,2},\psi_{T,1},\psi_{T,2})$ of the evaluation critics are iteratively updated by stochastic gradient descent (SGD) \cite{new21}, i.e.,
\begin{subequations}\label{eq:critic_update}
\begin{align}
\psi_{B,\ell} &\leftarrow \psi_{B,\ell} - \alpha_{\psi_B} \nabla_{\psi_{B,\ell}} \mathcal{L}(\psi_{B,\ell}), \ell =1,2 \label{eq:critic_update_B} \\
\psi_{T,\ell} &\leftarrow \psi_{T,\ell} - \alpha_{\psi_T} \nabla_{\psi_{T,\ell}} \mathcal{L}(\psi_{T,\ell}), \ell =1,2, \label{eq:critic_update_T}
\end{align}
\end{subequations}
where $\alpha_{\psi_B}$ and $\alpha_{\psi_T}$ denote the learning rates of the evaluation critics for Agent-B and Agent-T, respectively.

\subsubsection{Evaluation Actor Network Update} 
The evaluation actor network of Agent-B and Agent-T aim to obtain the optimal beamforming policy $\pi_B^*$ and trajectory policy $\pi_T^*$, respectively. Specifically, each actor maximizes the expected action value of its own first critic over the state distribution of the replay episodes, i.e,

\begin{subequations}\label{eq:optimal_policy}
\begin{align}
J_B(\phi_B)&\triangleq\mathbb{E}_{\mathbf{s}_B\sim\mathcal{D}}\Big[Q_B\big(n;\psi_{B,1}\big)\mid_{\mathbf{a}_B=\pi_B(\boldsymbol{\chi}_B[n];\phi_B)}\Big],\label{eq:optimal_policy_B}\\
J_T(\phi_T)&\triangleq\mathbb{E}_{\mathbf{s}_T\sim\mathcal{D}}\Big[Q_T\big(n;\psi_{T,1}\big)\mid_{\mathbf{a}_T=\pi_T(\boldsymbol{\chi}_T[n];\phi_T)}\Big].\label{eq:optimal_policy_T}
 \end{align}
\end{subequations}

Based on the deterministic policy gradient theorem\cite{new22}, the actor is updated by gradient ascent on \eqref{eq:optimal_policy}, where the policy gradient is evaluated using the first evaluation critic only (rather than the minimum of the twin critics), following TD3. In practice, the policy gradients are approximated over the mini-batch $\cal B$ as $\nabla_{\phi_B}J_B\approx\frac{1}{|\mathcal{B}|}\sum_{k=1}^{|\mathcal{B}|}\nabla_{\mathbf{a}}Q_B(n;\psi_{B,1})\mid_{\mathbf{a}=\pi_B(\boldsymbol{\chi}^{(k)}_B[n];\phi_B)}\;\nabla_{\phi_B}\pi_B(\boldsymbol{\chi}^{(k)}_B[n];\phi_B)$, and analogously for Agent-T with $J_T$, $\psi_{T,1}$, and $\boldsymbol{\chi}^{(k)}_T[n]$. The actor parameters are updated as
\begin{subequations}
    \begin{align}
        &\phi_B \leftarrow \phi_B + \alpha_{\phi_B} \nabla_{\phi_B} J_B(\phi_B)\\
        &\phi_T \leftarrow \phi_T + \alpha_{\phi_T} \nabla_{\phi_T} J_T(\phi_T),
    \end{align}
\end{subequations}
where $\alpha_{\phi_B}$ and $\alpha_{\phi_T}$ denote the learning rates of the evaluation actor networks of Agent-B and Agent-T, respectively. To stabilize the training process, each actor is updated once every $C$ critic updates, while the critics are updated at every iteration.

\subsubsection{Target Network Soft Update} 
Based on the soft target update mechanism \cite{new24}, the target actor-critic parameters of Agent-B and Agent-T are updated as 
\begin{subequations}\label{eq:soft_update}
\begin{align}
&\phi'_B \leftarrow \tau \phi_B + (1-\tau)\phi'_B \\
&\psi'_{B,\ell} \leftarrow \tau \psi_{B,\ell} + (1-\tau)\psi'_{B,\ell}, \ell=1,2\label{eq:soft_update_B} \\
& \phi'_T \leftarrow \tau \phi_T + (1-\tau)\phi'_T\\
&\psi'_{T,\ell} \leftarrow \tau \psi_{T,\ell} + (1-\tau)\psi'_{T,\ell}, \ell =1,2,\label{eq:soft_update_T}
\end{align}
\end{subequations}
 where $\tau\in(0,1) (\tau \ll 1)$ denotes the soft-update rate, which keeps the target networks slowly varying for stable training, and the target parameters $(\phi'_B,\phi'_T,\psi'_{B,\ell},\psi'_{T,\ell})$ are exactly those used in the TD targets. The soft update is performed once every $C$ critic updates, synchronized with the delayed actor update.

\subsection{Discussion of DeepDA-OPT}
Algorithm~\ref{alg:deepda} presents the proposed DeepDA-OPT algorithm for solving (P1), or equivalent (P0). The online interaction phase collects the episodes into the replay buffer $\cal D$ (Lines 4-14), while the training phase alternates between the critic, actor, and target network updates (Lines 16-25) until convergence.

% During the online experience interaction phase (Lines 4-14), Agent-B and Agent-T independently generate logical actions based on their local heterogeneous observations. Then, these logical outputs are mapped into the physical domain through the constrained noise exploration policy. Upon executing the joint physical actions, the environment computes the closed-form optimal receive beamformers via Proposition \ref{prop:optimal_filter} (Line 11). 

% After the episodic experiences are collected into the hierarchical experience replay buffer $\mathcal{D}$ (Line 15), the offline decentralized training phase is triggered (Lines 16-25). Then, a mini-batch experience set $\mathcal{B}$ is sampled per episode, and the twin evaluation critic networks of Agent-B and Agent-T are updated using the sequential TD targets to eliminate action-value overestimation. Finally, the evaluation actor networks and target networks of Agent-B and Agent-T are updated at a delayed frequency to ensure deep implicit collaboration and policy convergence without introducing explicit communication overhead.

\begin{algorithm}%[htbp]
\caption{Proposed DeepDA-OPT}
\label{alg:deepda}
\begin{algorithmic}[1]
\STATE \textbf{Initialize} eval-actor parameters $\phi_B, \phi_T$ and twin eval-critic parameters $\psi_{B\ell}, \psi_{T\ell}$ for $\ell \in \{1,2\}$.
\STATE \textbf{Initialize} target networks: $\phi'_B \leftarrow \phi_B$, $\phi'_T \leftarrow \phi_T$, $\psi'_{B\ell} \leftarrow \psi_{B\ell}$, $\psi'_{T\ell} \leftarrow \psi_{T\ell}$.
\STATE \textbf{Initialize} hierarchical experience replay buffer $\mathcal{D}$, and Ornstein-Uhlenbeck (OU) noise $\{\mathcal{N}_B, \mathcal{N}_T\}$.
\FOR{episode $= 1, 2, \ldots, E$}
    \STATE Observe the initial joint state $\mathbf{s}[0] = (\mathbf{s}_B[0], \mathbf{s}_T[0])$.
    \FOR{time slot $n = 0, 1, \ldots, N-1$}
        \STATE Generate logical actions $\tilde{\mathbf{a}}_B[n]$ and $\tilde{\mathbf{a}}_T[n]$ via eval-actors $\pi(\phi_B)$ and $\pi(\phi_T)$.
        \STATE Obtain physical ISAC transmit beamformers $\mathbf{a}_B[n]$ via the constrained noise exploration policy.
        \STATE Refine physical kinematics $\mathbf{a}_T[n]$ via the constrained noise exploration policy.
        \STATE Execute joint physical actions $(\mathbf{a}_B[n], \mathbf{a}_T[n])$ and observe next state $\mathbf{s}[n+1]$.
        \STATE Compute optimal receive beamformers $\mathbf{u}_{m,j}^*[n]$ and maximum sensing SINR $\lambda_{m,j}^{\max}[n]$ via Proposition \ref{prop:optimal_filter}.
        \STATE Evaluate the exact immediate reward $r[n]$ via \eqref{eq:reward}.
        \STATE Generate single-step experience tuple $\mathcal{T}[n]$.
    \ENDFOR
    \STATE Form complete episode sequence $\mathcal{E} = \{\mathcal{T}[n]\}_{n=0}^{N-1}$ to store into $\mathcal{D}$.
    \IF{offline training condition is met}
        \STATE Randomly sample a mini-batch experience set $\mathcal{B}$ from $\mathcal{D}$.
        \STATE Compute TD target values $y_B^{(k)}[n]$ and $y_T^{(k)}[n]$.
        \STATE Train eval-critics $\psi_{B\ell}$ and $\psi_{T\ell}$ by minimizing losses $\mathcal{L}_{B,\ell}$ and $\mathcal{L}_{T,\ell}$ via SGD.
        \IF{delayed policy update interval $d$ is met}
            \STATE Train eval-actors $\phi_B$ and $\phi_T$ via deterministic policy gradient ascent.
            \STATE Soft-update target networks $\phi'_B$, $\phi'_T$, $\psi'_{B\ell}$, and $\psi'_{T\ell}$ via \eqref{eq:soft_update}.
        \ENDIF
    \ENDIF
\ENDFOR
\end{algorithmic}
\end{algorithm}

\begin{table}%[t]
\centering
\caption{Comparison of Per-Slot Online Inference Complexity}
\label{tab:complexity_comparison}
\renewcommand{\arraystretch}{1.3} % 提升行高，增加排版的呼吸感
\footnotesize
\begin{tabular*}{\columnwidth}{@{} l @{\extracolsep{\fill}} l @{}}
\toprule
\textbf{Scheme} & \textbf{Per-Slot Online Inference Complexity} \\
\midrule
% 嵌套表格将方案名称断行，复杂度公式保持完整。
\begin{tabular}[t]{@{}l@{}}
DeepDA-OPT (ours) \\
\end{tabular} & $\mathcal{C}_{\mathrm{A}}(d_1,\mathbf{n}_1,a_1)+\mathcal{C}_{\mathrm{A}}(d_2,\mathbf{n}_2,a_2)+\mathcal{O}(MK_sS^3)$ \\
SA-TD3~\cite{ref14} & $\mathcal{C}_{\mathrm{A}}(d_{\mathrm{SA}},\bar{\mathbf{n}},a_{\mathrm{SA}})+\mathcal{O}(MK_sS^3)$ \\
SA-DDPG~\cite{ref23} & $\mathcal{C}_{\mathrm{A}}(d_{\mathrm{SA}},\bar{\mathbf{n}},a_{\mathrm{SA}})+\mathcal{O}(MK_sS^3)$ \\
DeepDA-MRC & $\mathcal{C}_{\mathrm{A}}(d_1,\mathbf{n}_1,a_1)+\mathcal{C}_{\mathrm{A}}(d_2,\mathbf{n}_2,a_2)+\mathcal{O}(MK_sS)$ \\
SA-TD3-MRC & $\mathcal{C}_{\mathrm{A}}(d_{\mathrm{SA}},\bar{\mathbf{n}},a_{\mathrm{SA}})+\mathcal{O}(MK_sS)$ \\
AO~\cite{ref06} & $\mathcal{O}\!\big(I_{\mathrm{AO}}[MK_sS^3+I_{\mathrm{b}}N_{\mathrm{b}}^3+I_{\mathrm{q}}N_{\mathrm{q}}^3]\big)$ \\
\bottomrule
\end{tabular*}
\end{table}

Table \ref{tab:complexity_comparison} presents the online inference complexities of various schemes for solving (P0). The action output dimensions of Agent-B and Agent-T are $a_1 = 2M(K_c+K_s)S$ and $a_2 = 2(K_c+K_s)$, respectively. The complexity of the dual-agent online forward propagation of DeepDA-OPT is $\mathcal{C}_{\mathrm{DA}}^{\mathrm{Actor}} = \mathcal{C}_{\mathrm{A}}(d_1,\mathbf{n}_1,a_1) + \mathcal{C}_{\mathrm{A}}(d_2,\mathbf{n}_2,a_2)$, where $\mathcal{C}_{\mathrm{A}}(d,\mathbf{n},a)
 = \mathcal{O}(3Dd + 3D^2 + Dn_1 + \sum_{l=1}^{L-1} n_l n_{l+1} + n_L a)$ denotes the forward-propagation complexity of one GRU-based actor with input dimension $d$, GRU hidden size $D$, $L$ hidden layers of width $\mathbf{n}=(n_1,...,n_L)$, and output dimension $a$. By default, the hidden-layer widths are set to $\mathbf{n}_1=(600,400)$ and $\mathbf{n}_2=(256,128)$. In addition, the environment side computes the optimal sensing receive beamformers of the $M$ BSs based on Proposition~\ref{prop:optimal_filter}, with a complexity of $\mathcal{O}(M K_s S^3)$.

\section{Numerical Results}
\label{sec:simulation}

In this section, we evaluate the performance of the proposed DeepDA-OPT scheme. Unless otherwise specified, the number of BSs is set to $M = 4$, each BS is equipped with $S = 6$ antennas (i.e., $2 \times 3$ UPA configuration), the number of C-UAVs/S-UAVs is set to $K_c = K_s=2$, and the aggregated sensing SINR threshold for target detection is $\Gamma_j = -10$ dB for $j\in{\cal K}_s$. The maximum allowable transmit power per BS is $P_{\max} = 10$ W (i.e., $40$ dBm), the carrier frequency is $f_c = 2.4$ GHz, and the antenna element spacing is $d = \lambda/2$. The flight altitude of all C-UAVs/S-UAVs is fixed at $H = 100$ m, and the maximum allowable flight speed is $V_{\max} = 20$ m/s. The number of flight time slots per episode is fixed to $N = 20$, and each time slot has a duration of $\Delta t = 1$~s. The path loss at the reference distance is set to $L_0 = -40$ dB, and the AWGN power is $\sigma_i^2 = 10^{-11}$ W (i.e., $-80$ dBm) for $i\in{\cal K}_c$. The noise variance at the sensing receiver is set to $\sigma_r^2=10^{-11}$ W (i.e., $-80$ dBm). The residual self-interference cancellation level of each BS is $\zeta_m = -130$ dB, and the inter-BS-interference suppression coefficient is $\eta_{l,m} = -50$ dB. The RCS variance of the C-UAVs/S-UAVs is set to $\sigma_t^2 = 1.0$.

The $M=4$ BSs are deployed at $(150, 300)$ m, $(200, 500)$ m, $(100, 100)$ m, and $(250, 200)$ m, respectively. The two C-UAVs depart from $(50, 150)$ m and $(50, 450)$ m, with destinations at $(350, 150)$ m and $(350, 450)$ m, respectively; the two S-UAVs depart from $(50, 250)$ m and $(50, 350)$ m, with destinations at $(350, 250)$ m and $(350, 350)$ m. The minimum collision-avoidance safety distance for all C-UAVs/S-UAVs is set to $d_{\min} = 10$ m.

The proposed DeepDA-OPT scheme is implemented in Python 3.8, with all DNNs built on PyTorch 1.12. The actor network of Agent-B adopts a deep MLP with hidden-layer sizes of $[600, 400]$, whereas that of Agent-T adopts a lightweight MLP with hidden-layer sizes of $[256, 128]$. For both agents, the twin critic evaluation networks adopt the same hidden-layer size as their corresponding actor networks, and each network further embeds a single GRU layer with a hidden dimension of $256$ after the input fully connected input layer. Table \ref{tab:hyperparameters} summarizes the key hyperparameters adopted in the offline training of the two agents.

\begin{table}[t]
\centering
\caption{Hyperparameter Settings for Proposed DeepDA-OPT Algorithm}
\label{tab:hyperparameters}
\renewcommand{\arraystretch}{1.2} 
\footnotesize %
\begin{tabular*}{\columnwidth}{@{} l @{\extracolsep{\fill}} l @{}}
\toprule
\textbf{Hyperparameter} & \textbf{Value} \\
\midrule
Total training episodes ($E$) & $60{,}000$ \\
Discount factor ($\gamma$) & $0.999$ \\
Replay buffer capacity ($\mathcal{D}$) & $2000$ \\
Soft update coefficient ($\tau$) & $10^{-4}$ \\
Mini-batch size ($|\mathcal{B}|$) & $64$ \\
Delayed update interval ($d$) & $2$ \\
Actor learning rate ($\alpha_{\phi_B}/\alpha_{\phi_T}$) & $2 \times 10^{-4}/1 \times 10^{-4}$ \\
Critic learning rate ($\alpha_{\psi_B}/\alpha_{\psi_T}$) & $5 \times 10^{-4}/5 \times 10^{-4}$ \\
OU noise std. dev. ($\sigma_B/\sigma_T$) & $0.2/0.25$ \\
OU mean reversion rate ($\theta_B/\theta_T$) & $0.2/0.2$ \\
Target policy noise std ($\tilde{\sigma}$) & $0.2$ \\
Target noise clip bound ($c$) & $0.5$ \\
Reward weights ($\omega_1, \dots, \omega_4$) & $10, 20, 15, 100$ \\
Urgency safety threshold ($\bar{\delta}^c/\bar{\delta}^s$) & $0.8/0.8$ \\
GRU hidden dim ($D$) / Layers & $256/1$ \\
Optimizer / Random seeds & Adam / $\{42, 24, 7, 123\}$ \\
\bottomrule
\end{tabular*}
\end{table}

To validate the effectiveness of the proposed DeepDA-OPT scheme, we consider the following five baseline schemes.

\begin{itemize} 
    \item \textit{AO Scheme \cite{ref06}:} This scheme employs an AO-based framework to iteratively solve (P1), in which the ISAC beamforming and C-UAV/S-UAV trajectory design subproblems are alternatively solved based on semidefinite relaxation (SDR) and SCA techniques.   
     \item \textit{SA-DDPG Scheme \cite{ref23}:} The classic single-agent DDPG algorithm is adopted, in which the heterogeneous transmit beamforming and trajectory actions are concatenated into a single joint action vector, and the sensing receive beamformers are obtained analytically as in Proposition~\ref{prop:optimal_filter}.    
   \item \textit{SA-TD3 Scheme \cite{ref14}:} A single centralized TD3 agent is employed, in which all heterogeneous actions are concatenated into a monolithic high-dimensional joint action vector, and the sensing receive beamformers are obtained analytically as in Proposition \ref{prop:optimal_filter}.
    \item \textit{DeepDA-MRC Scheme:} This scheme share the same dual-agent framework as DeepDA-OPT, but replaces the analytical receive beamformers in Proposition \ref{prop:optimal_filter} with the MRC receive beamformers $\mathbf{u}_{m,j} = \mathbf{a}_{m,j} / \|\mathbf{a}_{m,j}\|$, where $\mathbf{a}_{m,j}$ denotes the receive steering vector of S-UAV $j$ at the BS $m$. 
    \item \textit{SA-TD3-MRC Scheme:} This scheme uses the same single centralized TD3 architecture as SA-TD3, but replaces the analytical receive beamformers with the MRC receive beamformers.   
\end{itemize}

For fairness, all learning-based schemes adopt the same network architectures, hyperparameters, and training protocol as DeepDA-OPT, with only the input/output dimensions adapted to their respective action spaces. Note that all neural networks are trained with the Adam optimizer. Each learning-based scheme in four independent runs with global random seeds $\{42,24,7,123\}$, which govern all sources of run-level randomness, including network initialization, exploration noise, replay-buffer sampling, and controllable randomness in the environment. The final trained model of each run is evaluated over 100 episodes with exploration noise disabled, yielding 2,000 decision steps per seed and 8,000 decision steps per learning-based scheme. The aggregated sensing SINR is averaged over the S-UAVs and the evaluation time slots. A sensing outage occurs when the aggregated sensing SINR of any S-UAV falls below $\Gamma_j=-10$ dB in a time slot; a collision violation occurs when the distance between any UAV pair falls below $d_{\min}=10$ m.
Unless stated otherwise, the learning-based results and parameter sweeps are report as the cross-seed mean $\pm$ one sample standard deviation (SD).

\begin{figure}
\centering
\begin{minipage}{0.49\linewidth}
\centering
\includegraphics[width=0.9\linewidth]{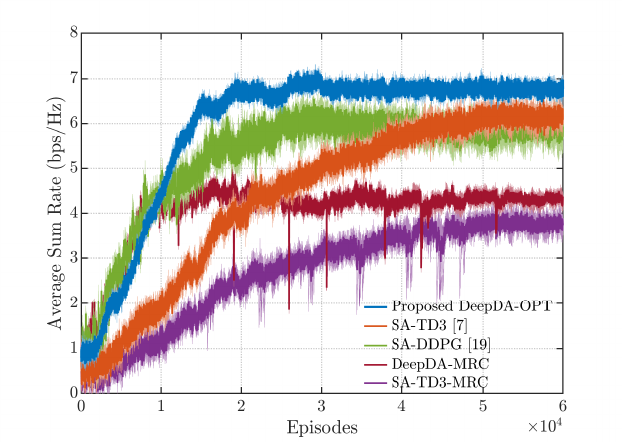}
\end{minipage}
\begin{minipage}{0.49\linewidth}
\centering
\includegraphics[width=0.9\linewidth]{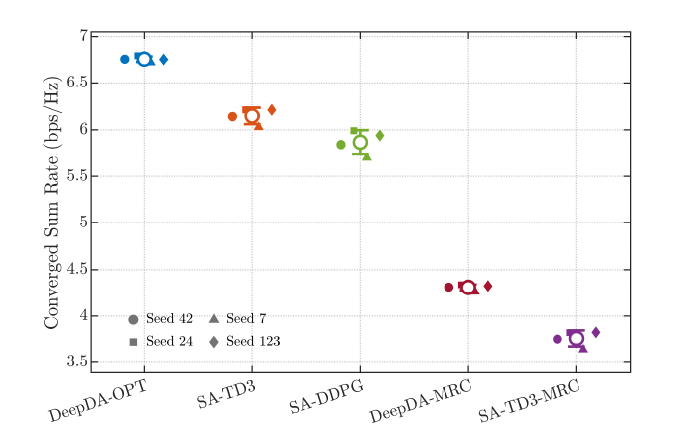}
\end{minipage}
  \caption{Training convergence and cross-seed converged sum-rate statistics of the learning-based schemes.}
  \label{fig:fig3}
\end{figure}

 Fig.~\ref{fig:fig3}(a) shows the training trajectories and their cross-seed variability, while Fig.~\ref{fig:fig3}(b) and Table~\ref{tab:ablation} compare the converged performance across the four runs. With the policy architecture fixed, replacing MRC with the analytical receive beamformer raises the converged sum rate from 4.31 to 6.76 bps/Hz for the dual-agent pair and from 3.76 to 6.15 bps/Hz for the single-agent pair. The analytical receiver maximizes the sensing SINR for given transmit beamformers and UAV positions by considering the interference-plus-noise covariance. Its rate benefit arises through the sensing-aware transmit-beamforming and trajectory decisions learned from this feedback.

 Table~\ref{tab:inference_latency} reports mean per-decision-step latency $\pm$ one sample SD, where the online computational latency of all learning-based schemes is measured on the same workstation (Intel Core i5-14600KF CPU and NVIDIA RTX 3090 GPU) under identical inference settings. The latency of each scheme is measured over the 100 evaluation episodes. AO requires 174.2$\pm$8.2~s to optimize one complete horizon-level instance with the default $N=20$; its latency is reported per instance because AO jointly optimizes the entire horizon. Under the default setting, DeepDA-OPT requires 2.15$\pm$0.46 ms per decision step on the CPU and 1.68$\pm$0.49 ms on the GPU, which is more than two orders of magnitude shorter than the 1~s slot duration and several orders of magnitude below the AO horizon-level solution time. These measurements substantiate the millisecond-level online execution capability of the proposed framework.

\begin{table}[t]
\centering
\caption{Measured online inference latency of the learning-based schemes.}
\label{tab:inference_latency}
\renewcommand{\arraystretch}{1.12}
\footnotesize
\begin{tabular}{lcc}
\toprule
{\textbf{Scheme}} & {\textbf{CPU latency (ms/step)}} & {\textbf{GPU latency (ms/step)}} \\
\midrule
{DeepDA-OPT(ours)} & {2.15$\pm$0.46} & {1.68$\pm$0.49} \\
{SA-TD3} & {1.25$\pm$0.27} & {1.14$\pm$0.42} \\
{SA-DDPG} & {1.24$\pm$0.24} & {1.03$\pm$0.34} \\
{DeepDA-MRC} & {1.86$\pm$0.32} & {1.69$\pm$0.55} \\
{SA-TD3-MRC} & {1.16$\pm$0.32} & {1.09$\pm$0.55} \\
\bottomrule
\end{tabular}
\end{table}

 \begin{table}[t]
\centering
\caption{Ablation of the analytical receive beamformer and dual-agent policy factorization.}
\label{tab:ablation}
\renewcommand{\arraystretch}{1.15}
\footnotesize
\setlength{\tabcolsep}{1.5pt}
\begin{tabular*}{\columnwidth}{@{} l @{\extracolsep{\fill}} c c c c c @{} }
\toprule
{\textbf{Scheme}} & {\shortstack{\textbf{Rx BF}\\\textbf{(Prop.~\ref{prop:optimal_filter})}}} & {\shortstack{\textbf{Dual}\\\textbf{agent}}} & {\shortstack{\textbf{Sum rate}\\\textbf{(bps/Hz)}}} & {\shortstack{\textbf{SD}\\\textbf{(bps/Hz)}}} & {\shortstack{\textbf{$E_{95}$}\\\textbf{(episodes)}}} \\
\midrule
{SA-TD3-MRC} & {$\times$} & {$\times$} & {3.76} & {0.09} & {41,815} \\
{DeepDA-MRC} & {$\times$} & {$\checkmark$} & {4.31} & {0.03} & {14,102} \\
{SA-TD3} & {$\checkmark$} & {$\times$} & {6.15} & {0.09} & {42,306} \\
{DeepDA-OPT (ours)} & {$\checkmark$} & {$\checkmark$} & {6.76} & {0.03} & {17,681} \\
\bottomrule
\end{tabular*}
\end{table}

 Table~\ref{tab:ablation} shows mean-rate gains of 0.61 bps/Hz with analytical reception and 0.55 bps/Hz with MRC from dual-agent factorization, where the receive-beamforming strategy is fixed. Cross-seed standard deviation (SD) also decreases for both receiver choices. Here, $E_{95}$ is the first episode when mean training performance reaches $95\%$ of its own converged level and remains above it. Factorization reduces $E_{95}$ from 42,306 to 17,681 episodes with analytical reception and from 41,815 to 14,102 episodes with MRC. These matched comparisons support more efficient and consistent learning of the heterogeneous beamforming and mobility mappings coordinated through the shared reward.

Table \ref{tab:constraint_statistics} reports cross-seed means and one sample SD, with AO evaluated over $10$ independent optimization instances. DeepDA-OPT's mean aggregated sensing SINR is $-5.60\pm0.10$ dB, i.e., an average margin of approximately $4.40$ dB above the required threshold, with a sensing-outage probability of $0.25\pm0.14\%$. Evaluated trajectories show no collision violation for all the schemes, and the minimum inter-UAV distance is $43.1\pm2.7$ m.

\begin{table*}[t]
\centering
\caption{Constraint-satisfaction statistics under the default setting.}
\label{tab:constraint_statistics}
\renewcommand{\arraystretch}{1.15}
\footnotesize
\begin{tabular*}{\textwidth}{@{} l @{\extracolsep{\fill}} c c c c @{} }
\toprule
{\textbf{Scheme}} & {\begin{tabular}[c]{@{}c@{}}\textbf{Mean aggregated}\\\textbf{sensing SINR (dB)}\end{tabular}} & {\begin{tabular}[c]{@{}c@{}}\textbf{Sensing outage}\\\textbf{(\%)}\end{tabular}} & {\begin{tabular}[c]{@{}c@{}}\textbf{Collision violation}\\\textbf{(\%)}\end{tabular}} & {\begin{tabular}[c]{@{}c@{}}\textbf{Minimum inter-UAV}\\\textbf{distance (m)}\end{tabular}} \\
\midrule
{DeepDA-OPT (ours)} & {$-5.60 \pm 0.10$} & {$0.25 \pm 0.14$} & {$0.00 \pm 0.00$} & {$43.1 \pm 2.7$} \\
{SA-TD3} & {$-7.17 \pm 0.18$} & {$2.26 \pm 0.28$} & {$0.00 \pm 0.00$} & {$55.0 \pm 4.4$} \\
{SA-DDPG} & {$-7.25 \pm 0.20$} & {$3.43 \pm 0.50$} & {$0.00 \pm 0.00$} & {$69.4 \pm 5.7$} \\
{DeepDA-MRC} & {$-7.52 \pm 0.11$} & {$3.84 \pm 0.15$} & {$0.00 \pm 0.00$} & {$64.0 \pm 3.0$} \\
{SA-TD3-MRC} & {$-8.05 \pm 0.20$} & {$4.71 \pm 0.45$} & {$0.00 \pm 0.00$} & {$69.8 \pm 6.5$} \\
{AO} & {$-6.60$} & {$0.00$} & {$0.00$} & {$70.5$} \\
\bottomrule
\end{tabular*}
\end{table*}

\begin{figure}[htbp]
  \centering
\includegraphics[width=0.65\linewidth]{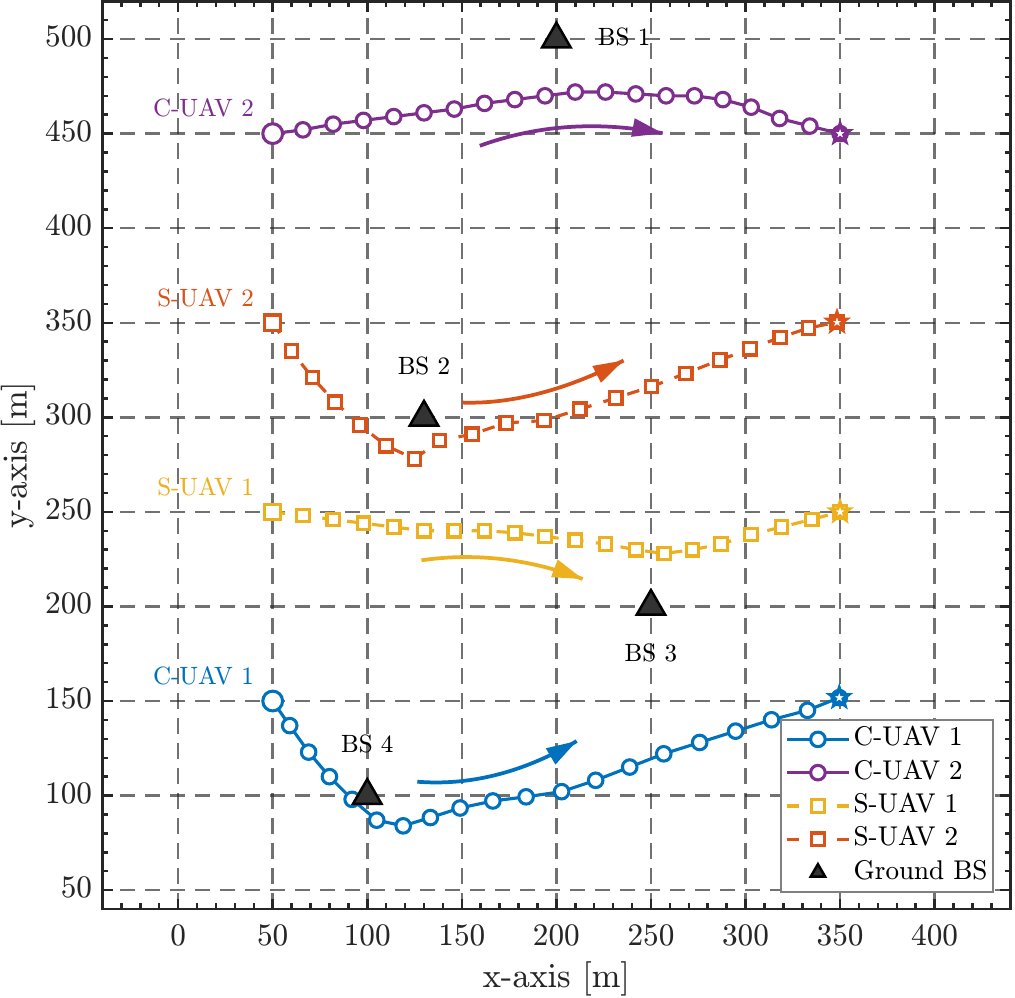}
  \caption{UAV flight trajectories by the proposed DeepDA-OPT algorithm with $K_c=K_s=2$ and $M=4$.}
  \label{fig:fig4}
\end{figure}

Fig. \ref{fig:fig4} shows the UAV flight trajectories in the horizontal $x-y$ plane under DeepDA-OPT with $K_c=K_s=2$ and $M=4$. The trajectories of the C-UAVs exhibit channel-aware maneuvering toward the BSs. C-UAV~1 first deviates downward to approach BS~4 and then passes through the coverage area of BS~3, while C-UAV~2 tends to approach BS~1. The S-UAVs maintain smooth trajectories while keeping safe distances from the C-UAVs, and no trajectory crossings or collisions occur over the entire operation horizon. This indicates that the collision penalty $P_{col}[n]$ in \eqref{eq:reward} effectively guides Agent-T to learn safe policies.

\begin{figure}[htbp]
\centering
\includegraphics[width=0.65\linewidth]{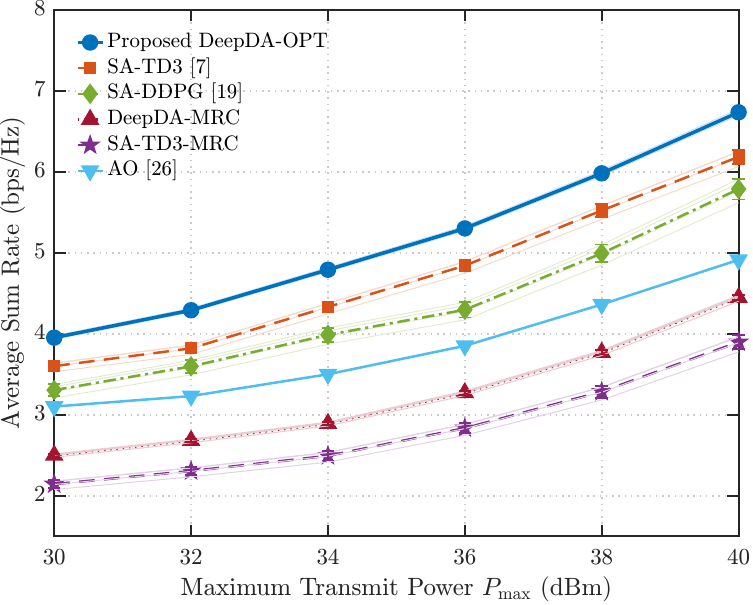}
\caption{The average sum rate versus BS transmit power budget $P_{\max}$.}
\label{fig:fig5}
\end{figure}

Fig.~\ref{fig:fig5} shows the average sum rate versus the BS transmit power budget. DeepDA-OPT retains the highest average sum rate across the tested power budgets, and its advantage over SA-TD3 grows from 0.35~bps/Hz at 30~dBm to 0.61~bps/Hz at 40~dBm. For baselines, AO increases from 3.10 to 4.91~bps/Hz and remains above both MRC-based variants. The cross-seed SD of DeepDA-OPT stays within 0.018–0.03~bps/Hz across the sweep, below that of SA-TD3 (0.05–0.09~bps/Hz) and SA-DDPG (0.07–0.13~bps/Hz), indicating consistent learning through policy factorization. The gain stems from the division of labor, where Agent-B specializes in the multi-BS transmit-beamforming policy, and Agent-T adjusts the UAV geometry under the shared reward, so that neither actor has to represent both types of control.

\begin{figure}[htbp]
\centering
\includegraphics[width=0.65\linewidth]{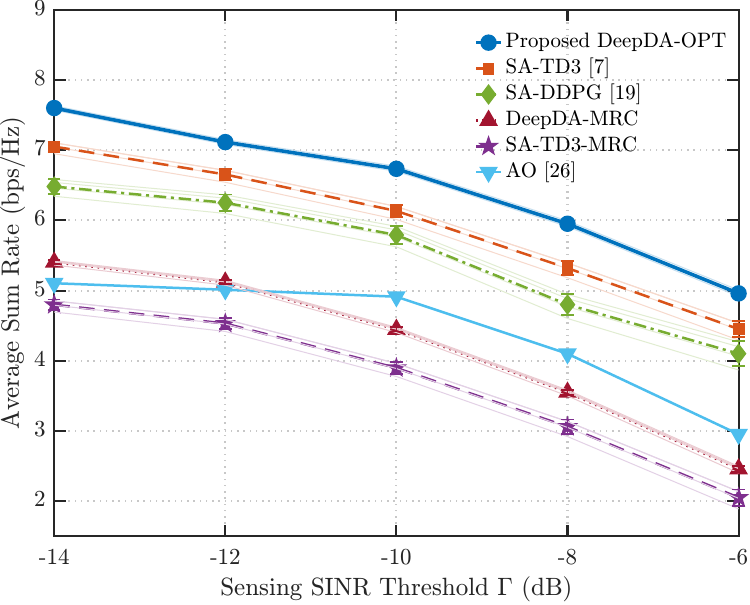}
\caption{Average sum rate of C-UAVs versus sensing SINR threshold $\Gamma$.}
\label{fig:fig6}
\end{figure}

Fig.~\ref{fig:fig6} shows the average sum rate versus the required sensing SINR threshold. As the threshold increases, the average sum rate decreases for all schemes, since stronger sensing requirements intensify the competition for transmit power and spatial degrees of freedom. DeepDA-OPT achieves the highest average sum rate, exceeding SA-TD3 by 0.46--0.63~bps/Hz, with the smallest cross-seed SD (0.03--0.04~bps/Hz, versus 0.07--0.11 for SA-TD3 and 0.11--0.18 for SA-DDPG), indicating more consistent learning. The gain of DeepDA-OPT stems from the division of labor: Agent-B optimizes the communication–sensing beam allocation via the analytical sensing feedback, while Agent-T relocates the UAV to reshape the channel and interference geometry.

\begin{figure}[htbp]
  \centering
  \includegraphics[width=0.65\linewidth]{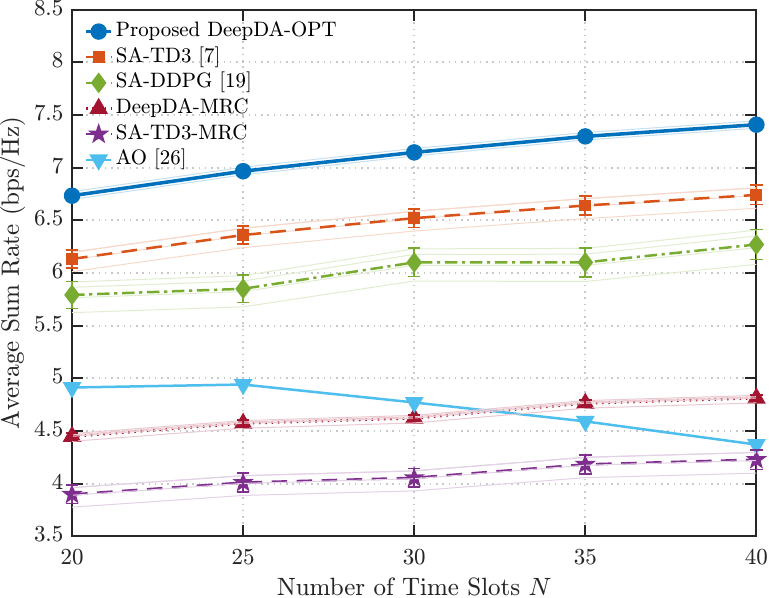}
  \caption{The average sum rate of C-UAVs versus the flight horizon $N$.}
  \label{fig:fig7}
\end{figure}

Fig.~\ref{fig:fig7} shows the average sum rate versus the flight horizon $N$. The average sum rates of the learning-based schemes increase with $N$, with the advantage over SA-TD3 remaining within 0.6~bps/Hz. For the baselines, the rate of AO is non-monotonic, while DeepDA-MRC remains below AO at $N\in\{20,25,30\}$ but exceeds it at $N\in\{35,40\}$. The SD of DeepDA-OPT stays near 0.03~bps/Hz, that of SA-TD3 near 0.09~bps/Hz, and that of SA-DDPG rises from 0.13 to 0.14~bps/Hz, indicating more consistent learning as the temporal coupling strengthens with longer horizons. The persistent gain stems from the spatial–temporal coordination: a longer horizon gives Agent-T more freedom to allocate flight time to favorable channel geometries, and Agent-B adapts the transmit beams to the resulting positions.
 
\begin{figure}[htbp]
  \centering
  \subfloat[Versus the number of C-UAVs $K_c$ \label{fig:fig8}]{
    \includegraphics[width=0.4\linewidth]{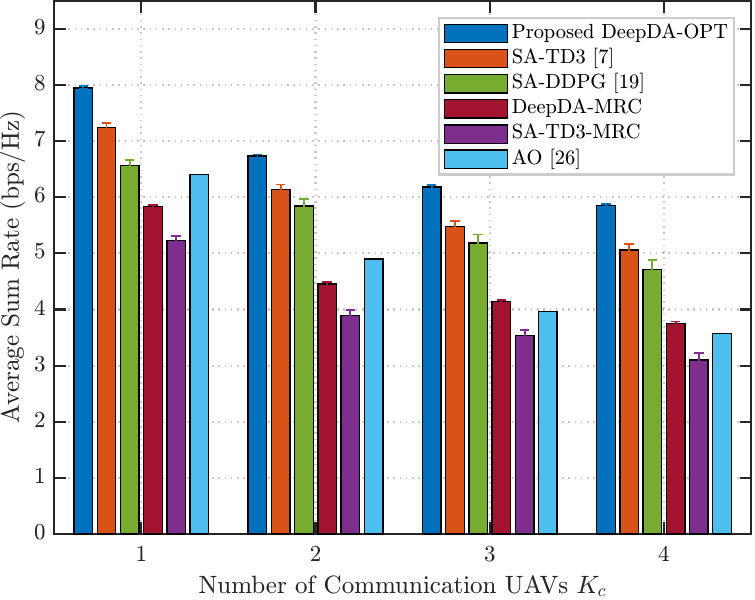}
  }\hfill
  \subfloat[Versus the number of S-UAVs $K_s$ \label{fig:fig9}]{
    \includegraphics[width=0.4\linewidth]{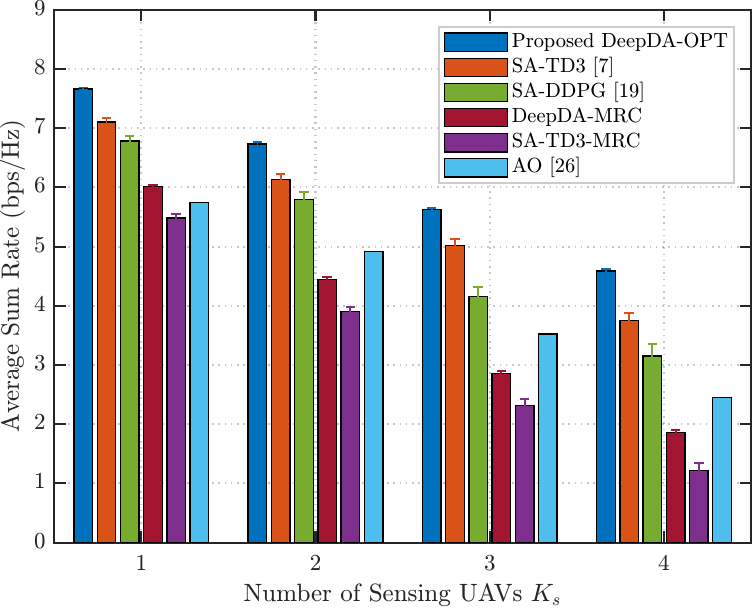}
  }
  \caption{The average sum rate with different $K_c$ and $K_s$.}
  \label{fig:num_uavs}
\end{figure}

Fig.~\ref{fig:fig8} and Fig.~\ref{fig:fig9} show the average sum rate versus the numbers of C-UAVs and S-UAVs, respectively. In Fig.~\ref{fig:fig8}, increasing $K_c$ from 1 to 4 adds multiplexed communication streams and strengthens the multiuser interference under fixed BS power budgets; in Fig.~\ref{fig:fig9}, increasing $K_s$ adds sensing constraints and multi-target interference, intensifying the competition with communication for power and spatial resources. Hence, the average sum rates of all schemes decrease. DeepDA-OPT achieves the highest average sum rate in both sweeps, exceeding SA-TD3 by 0.61--0.79~bps/Hz and 0.56--0.83~bps/Hz, respectively.
The SDs of DeepDA-OPT increase moderately from 0.03 to 0.04~bps/Hz and from 0.02 to 0.04~bps/Hz across the two sweeps, below those of SA-TD3 (0.07--0.11 and 0.06--0.13~bps/Hz) and SA-DDPG (0.10--0.18 and 0.09–0.20~bps/Hz), indicating more consistent learning as the action space expands.

 To evaluate robustness to imperfect downlink CSI, we perturb the channel observations available to Agent-B while retaining the underlying channel model and the other simulation settings. Conditional on the corresponding physical channel, the observed channels are modeled as
\begin{subequations}\label{eq:csi_perturbation}
\begin{align}
\widehat{\mathbf{h}}_{m,i}^{c}[n]
 = \mathbf{h}_{m,i}^{c}[n]+\mathbf{e}_{m,i}^{c}[n],\\
\widehat{\mathbf{h}}_{m,j}^{s}[n]
 = \mathbf{h}_{m,j}^{s}[n]+\mathbf{e}_{m,j}^{s}[n]
\end{align}
\end{subequations}
 where $\mathbf{e}_{m,i}^{c}[n] \sim\mathcal{CN}(\mathbf{0},
 \rho\|\mathbf{h}_{m,i}^{c}[n]\|_2^2\mathbf{I}_S)$, $\mathbf{e}_{m,j}^{s}[n]
 \sim\mathcal{CN}(\mathbf{0},\rho\|\mathbf{h}_{m,j}^{s}[n]\|_2^2\mathbf{I}_S)$, and $\rho\geq0$ is a CSI-error coefficient. The error vectors are drawn independently across links and time slots. Since each error entry has variance $\rho\|\mathbf{h}\|_2^2$, the conditional normalized mean-squared error (NMSE) of the channel vector is $S\rho$; with the default $S=6$, $\rho=0.1$ corresponds to an NMSE of $0.6$, whereas $\rho=0$ corresponds to perfect CSI. The true channels are used to evaluated the resulting ISAC performance, whereas Agent-B acts on the perturbed channel observations. We retain the four-seed evaluation protocol and reporting convention above.

\begin{figure}[htbp]
  \centering
  \IfFileExists{fig10.pdf}{
    \includegraphics[width=0.65\linewidth]{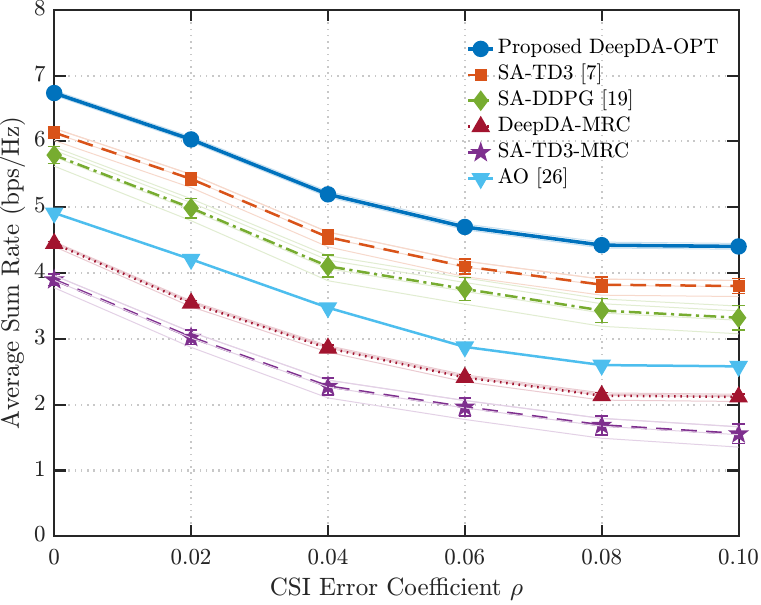}%
  }{ \fbox{\parbox[c][1.8in][c]{0.88\linewidth}{\centering\color{red}
      Insert the CSI-robustness plot here.\\
      Save the figure as \texttt{fig10.pdf} (width: 0.9\textbackslash linewidth).}}
  }
  \caption{Average sum rate versus the CSI-error coefficient $\rho$.}
  \label{fig:csi_robustness}
\end{figure}

 Fig.~\ref{fig:csi_robustness} shows the average sum rate versus the CSI-error coefficient $\rho$. DeepDA-OPT achieves $6.76\pm0.03$~bps/Hz at $\rho=0$; as $\rho$ increases 0.1, its rate falls to $4.40\pm0.04$~bps/Hz. Noisy channel observations degrade the beam-direction selection and interference suppression of Agent-B, and coordinated trajectory control cannot fully compensate for this mismatch. Nevertheless, DeepDA-OPT exceeds all baselines across all the sweep: at $\rho=0.1$, SA-TD3 and AO achieve 3.8 and 2.58~bps/Hz, respectively, compared with 2.12~bps/Hz for DeepDA-MRC. The benefit of the matched analytical receiver is 2.3~bps/Hz for DeepDA-OPT over DeepDA-MRC and 2.25~bps/Hz for SA-TD3 over its MRC-based counterpart. This spread remains well below that of SA-TD3 (0.12~bps/Hz) and SA-DDPG(0.19~bps/Hz) at $\rho=0.1$, indicating that the specialized beamforming and mobility policies maintain stable learned solutions despite the rate loss.

\begin{figure}[t]
  \centering
  \subfloat[Varying $M$ ($S=6$).\label{fig:scalability_M}]{%
    \IfFileExists{fig11a.pdf}{%
      \includegraphics[width=0.48\linewidth]{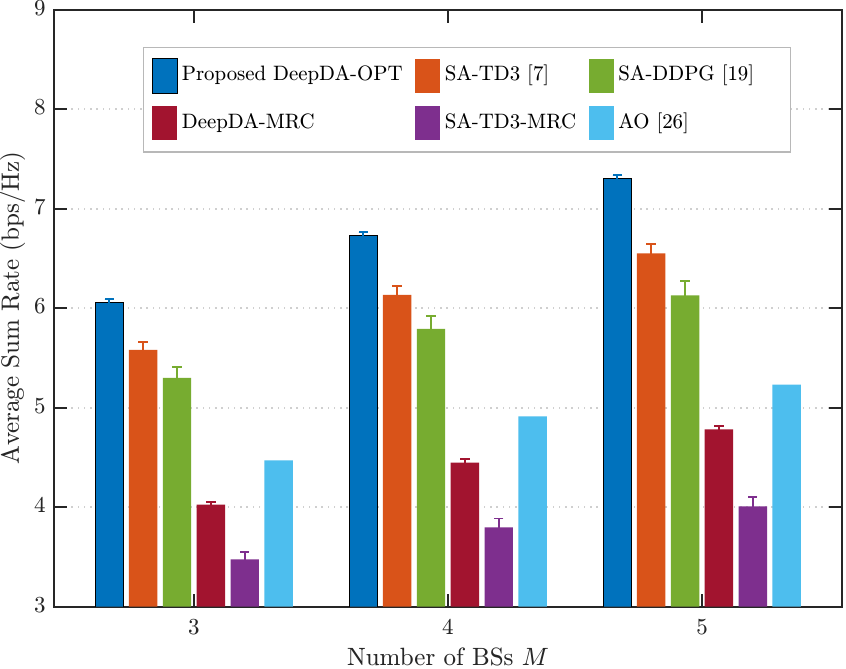}%
    }{%
      \fbox{\parbox[c][1.25in][c]{0.46\linewidth}{\centering\color{red}
        Insert panel (a) here.\\Save as \texttt{fig11a.pdf}.}}
    }%
  }\hfill
  \subfloat[Varying $S$ ($M=4$).\label{fig:scalability_S}]{%
    \IfFileExists{fig11b.pdf}{%
      \includegraphics[width=0.48\linewidth]{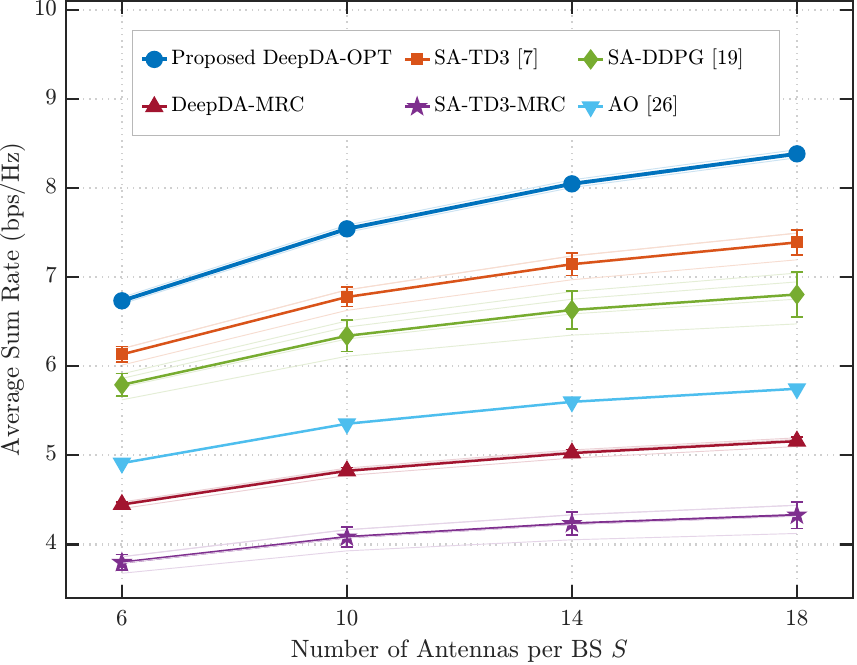}%
    }{%
      \fbox{\parbox[c][1.25in][c]{0.46\linewidth}{\centering\color{red}
        Insert panel (b) here.\\Save as \texttt{fig11b.pdf}.}}
    }%
  }\par\vspace{1mm}
  \subfloat[Joint $(M,S)$ sweep for DeepDA-OPT.\label{fig:scalability_joint}]{%
    \IfFileExists{fig11c.pdf}{%
      \includegraphics[width=0.90\linewidth]{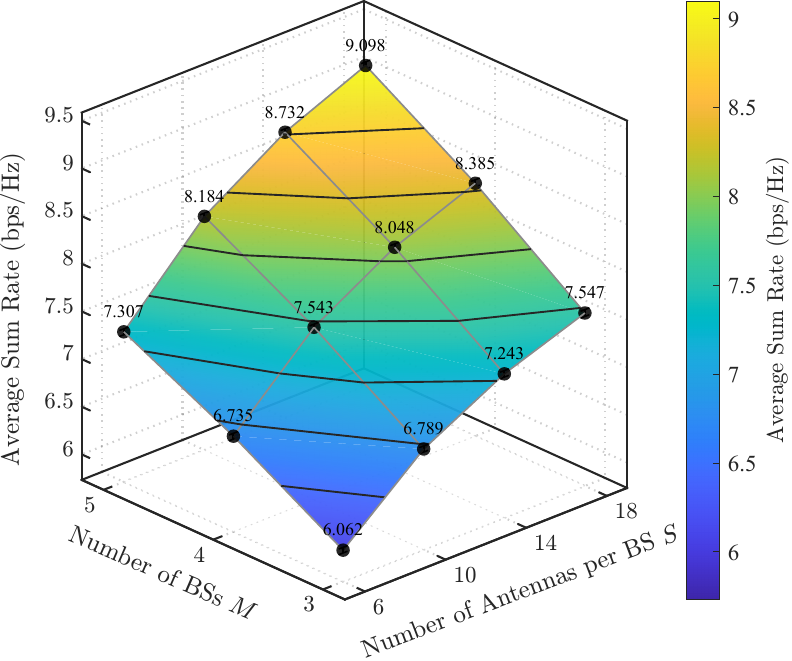}%
    }{%
      \fbox{\parbox[c][1.85in][c]{0.7\linewidth}{\centering\color{red}
        Insert panel (c) here.\\Save as \texttt{fig11c.pdf}.}}
    }%
  }
  \caption{Scalability of the average sum rate with the number of BSs and antennas per BS.}
  \label{fig:scalability}
\end{figure}

 Fig.~\ref{fig:scalability_M} shows the average sum rate versus the number of BSs at $S=6$. DeepDA-OPT achieves $6.06\pm0.03$, $6.76\pm0.03$, and $7.31\pm0.03$ bps/Hz as $M$ increases from 3 to 5, since adding BSs with fixed per-BS power budgets enlarges the aggregate transmit power and the distributed spatial diversity. AO improves from 4.47 to 5.23 bps/Hz and remains above both MRC-based variants. SA-TD3 achieves $5.58\pm0.08$, $6.15\pm0.09$, and $6.55\pm0.10$ bps/Hz, so the advantage of DeepDA-OPT grows from 0.48 to 0.76~bps/Hz while its SD stays at 0.03~bps/Hz. This indicates that the factorized policy exploit the additional BS resources even as the transmit-beamforming action dimension growing with $M$.

 Fig.~\ref{fig:scalability_S} shows the average sum rate versus the antenna number per BS at $M=4$. DeepDA-OPT rises from $6.76\pm0.03$ bps/Hz at $S=6$ to $8.39\pm0.04$ bps/Hz at $S=18$, and SA-TD3 rises from $6.15\pm0.09$ to $7.39\pm0.14$ bps/Hz, as more antennas provide array gain and additional spatial degrees of freedoms for shaping desired beams and suppressing interference. AO increases from 4.91 to 5.75 bps/Hz and remains above both MRC-based variants. The advantage of DeepDA-OPT over SA-TD3 grows from 0.61 to 1~bps/Hz. The SD of DeepDA-OPT changes only from 0.03 to 0.04~bps/Hz, compared with 0.09–0.14 for SA-TD3 and 0.13–0.25 for SA-DDPG, showing that the specialized beamforming policy scales more consistently than a single actor representing both types of control.

 Fig.~\ref{fig:scalability_joint} shows the joint sweep of DeepDA-OPT over $M$ and $S$. The average sum rate increases from $6.06\pm0.03$~bps/Hz at $(M,S)=(3,6)$ to $9.10\pm0.04$~bps/Hz at $(M,S)=(5,18)$. For $K_c=K_s=2$, the real-valued transmit-beamforming action dimension $2M(K_c+K_s)S$ grows from 144 to 720 between these endpoints. The improved rate and the modest change in SD indicate that the division of labor maintains effective learning as the action space expands. Figs.~\ref{fig:scalability_M} and \ref{fig:scalability_S} compare DeepDA-OPT with the baselines, whereas Fig.~\ref{fig:scalability_joint} reports DeepDA-OPT only. These results indicate scalability of DeepDA-OPT within the tested configurations.

\section{Conclusion}
\label{sec:conclusion}
 In this paper, we investigated the joint multi-BS multi-UAV ISAC beamforming and trajectory design with the objective of maximizing the average sum rate of C-UAVs over the entire flight episode, subject to the BS transmit power and sensing SINR constraints of S-UAVs, as well as C-UAV/S-UAV flight mission and collision avoidance constraints. First, we obtained the optimal BS receive beamformers in closed form based on the generalized Rayleigh quotient theorem. Next, the joint design problem was reformulated as a constrained MDP, and we developed a dual-agent DRL-based DeepDA-OPT scheme, where Agent-B and Agent-T are trained based on the TD3 policy gradient algorithm for learning the multi-BS transmit beamforming design and C-UAVs/S-UAVs trajectory design, respectively. 
 The numerical results under the evaluated configurations demonstrate that, compared with the baselines, DeepDA-OPT offers a favorable tradeoff among communication performance, empirical constraint satisfaction, training behavior, and online execution latency, while retaining its performance advantage across different numbers of BSs.
  
 Note that in order to enhance the robustness of the proposed dual-agent DRL-based DeepDA-OPT scheme, it is worthwhile to investigate several practical UAV-ISAC issues, such as stochastic RCS models for sensing, correlated Doppler shifts/channel-aging processes induced by the high mobility of UAVs, and persistent blockage/NLoS propagation arising from urban or terrain environments, which are left for future work. In addition, by capturing the UAV flight dynamics, an interesting extension is to incorporate UAV propulsion-energy constraints into the joint beamforming and trajectory design framework, where the propulsion power consumption is closely coupled with the flight attitude, velocity, and acceleration. This extension is particularly relevant for mission-critical or energy-efficient ISAC scenarios, such as emergency rescue and infrastructure inspection.

\bibliographystyle{IEEEtran}
\bibliography{ref}

\end{document}